\documentstyle[11pt]{article}
\begin{document}
\begin{titlepage}

\vspace{0.2cm}

\title{ Top-quark pair production via polarized and unpolarized protons
          in the supersymmetric QCD \footnote{Supported in part by Committee of
         National Natural Science Foundation of China and Project IV.B.12 of
         scientific and technological cooperation agreement between China and
         Austria}}

\author{{ \ Yu Zeng-Hui $^{a,c}$ \ Herbert Pietschmann $^{a}$
\ Ma Wen-Gan  $^{b,c}$ \ Han Liang  $^{c}$  \ Jiang Yi  $^{c}$
 }\\\
{\small $^{a}$Institut f\" ur Theoretische Physik, Universit\" at Wien, A-1090 Vienna, Austria} \\
{\small $^{b}$CCAST (World Laboratory), P.O.Box 8730, Beijing 100080,P.R.China} \\
{\small $^{c}$Department of Modern Physics, University of Science and Technology}\\
{\small of China (USTC), Hefei, Anhui 230027, P.R.China}\\
}
\date{}
\maketitle

\vskip 12mm

\begin{center}\begin{minipage}{5in}

\begin{center} ABSTRACT\end{center}
\baselineskip 0.3in

{The QCD corrections to the top-quark pair production via
both polarized and unpolarized gluon fusion in $pp$ collisions
are calculated in the Minimal Supersymmetric Model(MSSM).
We find the MSSM QCD corrections can reach $4 \%$ and may be observable
in future precise experiments. Furthermore,
we studied the CP violation in the
MSSM, our results show that the CP violating parameter is sensitive to
the masses of SUSY particles (It becomes zero, when the c.m. energy is
less than twice the masses of both gluino and stop quarks.)
and may reach $10^{-3}$.} \\

\vskip 10mm

{~~~~PACS number(s): 13.65.+i, 13.88.+e, 14.65.-q, 14.80.Dq, 14.80.Gt}
\end{minipage}
\end{center}
\end{titlepage}

\baselineskip=0.36in

\eject
\rm
\baselineskip=0.36in

\begin{flushleft} {\bf I. Introduction} \end{flushleft}
\par
  The minimal supersymmetric model(MSSM) \cite{s1} is one of the most
interesting extensions of the Standard Model (SM). Therefore
testing the MSSM has attracted much interest. As is well known, the
MSSM predicts supersymmetric(SUSY) partners to all particles expected
by the SM, and searching for their existence is very important.
\par
  Since the top-quark was already found experimentally by the CDF and D0
Collaborations at Fermilab \cite{s2}, we believe that more and more
experimental events including top-quark will be collected in future experiments.
 That gives us a good chance to study the physics in top-quark pair
production from $pp$ or $p \bar p$ collisions with more precise experimental
results. Because of the heavy mass of the top quark this process provides
a test of the SM and possible signals of new physics at high energy.
\par
The dominant subprocesses of top-quark pair production in $pp$ or $p\bar p$
colliders are quark-antiquark annihilation and gluon-gluon
fusion. The lowest order of those two subprocesses has been studied in
Ref. \cite{s3}. There it was found that the former subprocess ($q\bar q$
annihilation) is more dominant in $p\bar p$ collisions when the c.m.
energy($\sqrt{s}$) is near the threshold value $2 m_{t}$, whereas subprocess
via $gg$ fusion will be more and more important with increasing c.m.
energy, and can become the most dominant one when the c.m. energy is
much larger than $2 m_{t}$.

\par
In Ref. \cite{s4}, the QCD corrections to top-quark pair production in
$p\bar p$ collisions have been studied in the frame of the SM. It may
seem natural that the QCD corrections of those processes in the frame of
the MSSM are important for distinguishing those two models. Recently,
the SUSY QCD corrections to top pair production via $q \bar q$ annihilation
were given in Ref. \cite{s5}. The SUSY QCD corrections via unpolarized
gluon-gluon fusion were presented by C.S.Li. et. al \cite{s6}.
\par
It is obvious that the correction from the SUSY QCD is related to
the masses of top-quark and SUSY particles. Assuming the SUSY breaking
scale at about 1 TeV, the masses of SUSY particles would be smaller
than 1 TeV. So we can hope that corrections from SUSY particles are
significant, since the heavy mass of the top quark ($m_{t}= ~175.6\pm
5.5 ~ GeV$ (world average)) may be comparable to some of the light
SUSY particle masses. Therefore the SUSY QCD correction would give
us some significant information about the existence of SUSY particles
indirectly.
\par
Recently, the spin structure of the nucleon has been intensely studied
by polarized deep inelastic scattering experiments at CERN and SLAC.
This knowledge allows us to find a clear signal beyond the SM,
if we collect enough events in the process of top-quark pair production
from polarized $pp$ or $p \bar{p}$ collisions. In the SM QCD, there is
no CP violation mechanism, whereas in the SUSY QCD, the situation may
be different. If we introduce phase angle of quark SUSY
partners, we can get CP violation in the MSSM QCD \cite{s7}. Once we
get enough statistics of top-quark pairs from $pp$ or $p \bar{p}$ colliders
at higher energy, it will be possible to test CP violation. On the other hand,
the spin-dependent parton distributions can be obtained from their polarized
structure function data in Ref.\cite{Geh}\cite{GLK}\cite{STR}.
There one
found
that the shape of polarized gluon and quark distributions
in the nucleon depends on its polarization.
Therefore the CP violation effects through the process of
top-quark pair production via $gg$ fusion may be observed
in polarized $pp$ or $p \bar{p}$ collisions.

\par
In this work we concentrate on the SUSY QCD corrections to the process
$pp \rightarrow g g \rightarrow t \bar{t} X$ both in polarized and unpolarized
colliding beams. In section 2, we give the tree level contribution to
subprocess $g g \rightarrow t \bar{t}$. In section 3 we give the analytical
expressions of the SUSY QCD corrections to $gg \rightarrow t \bar{t}$.
In section 4 the numerical results of the subprocess $gg \rightarrow
t \bar{t}$ and the process $pp \rightarrow g g \rightarrow t \bar{t} X$
are presented. The conclusion is given in section 5 and some details of
the expressions are listed in the appendix.

\begin{flushleft} {\bf II. The Tree-Level Subprocess} \end{flushleft}
\par
The graphical representation of the process $g(\lambda_{1},k_{1})
g(\lambda_{2},k_{2}) \rightarrow t(p_{1}) \bar{t} (p_{2})$ is shown
in Fig.1 (a). The Mandelstam variables are defined as usual
$$
\begin{array} {lll}
    \hat{s} & =(p_{1}+p_{2})^2=(k_{1}+k_{2})^2
\end{array}
\eqno {(2.1)}
$$
$$
\begin{array} {lll}
    \hat{t} & =(p_{1}-k_{1})^2=(k_{2}-p_{2})^2
\end{array}
\eqno {(2.2)}
$$
$$
\begin{array} {lll}
   \hat{u} & =(p_{1}-k_{2})^2=(k_{1}-p_{2})^2
\end{array}
\eqno {(2.3)}
$$
 so $\hat{s}+\hat{t}+\hat{u}=2 m_{t}^2$.
The amplitude of tree-level diagrams with polarized gluons can be written
as:\cite{s3} ($a, b$ are color indices of external gluons , $i, j$ are
colors of external top-quarks and $T^{a}=\frac{\lambda_{a}}{2}$ are the
Gell-Mann matrices.)
$$
\begin{array} {lll}
M_{0}^{(l)} &~ = ~ g_{s}^2 \epsilon ^{\mu,a} (\lambda_{1},k_{1}) \epsilon ^{\nu,b} (\lambda_{2},k_{2}) \bar{u}_{i}(p_{1})
\Gamma^{(l)} v_{j}(p_{2}), ~~~(l=s,t,u) \\
\end{array}
\eqno {(2.4)}
$$
with
$$
\begin{array} {lll}
  \Gamma^{(s)} &~= ~ \frac{T^{c}_{ij} f_{abc}}{s} [(\rlap/k_{1} - \rlap/k_{2}) g_{\mu\nu}+
  (2 k_{2} + k_{1})_{\mu} \gamma_{\nu} - (2 k_{1} + k_{2})_{\nu} \gamma_{\mu}]
\end{array}
\eqno {(2.5)}
$$
$$
\begin{array} {lll}
  \Gamma^{(t)} &~= ~ \frac{-i T_{im}^{a} T_{mj}^{b}}{t-m_{t}^{2}} \gamma _{\mu}
  (\rlap/k_{2}-\rlap/p_{2}+m_{t})\gamma _{\nu }
\end{array}
\eqno {(2.6)}
$$
$$
\begin{array} {lll}
 \Gamma^{(u)} &~= ~ \frac{-i T_{im}^{b} T_{mj}^{a}}{u-m_{t}^{2}}\gamma _{\nu }(\rlap/k_{1}-\rlap/p_{2}+m_{t})
 \gamma _{\mu }
\end{array}
\eqno {(2.7)}
$$
 We chose a form in which only physical polarizations of gluons remained:
$$
\begin{array} {lll}
\epsilon^{\mu \ast} (\lambda_{1},k_{i}) \epsilon^{\nu} (\lambda_{2},k_{i})~=~
\frac{\delta_{\lambda_1,\lambda_2}}{2} (-g^{\mu\nu}+
\frac{n^{\mu} k_{i}^{\nu}+n^{\nu} k_{i}^{\mu}}{n \cdot k_{i}}-
\frac{n^2 k_{i}^{\mu} k_{i}^{\nu}}{(n \cdot k_{i})^2} +
i \lambda_1 \epsilon^{\sigma \mu \rho \nu} \frac{k_{i \sigma} n_{\rho}}
{n\cdot k_{i}}).
\end{array}
\eqno {(2.8)}
$$
where $n=k_1+k_2$, $\lambda_{1,2}= \pm 1$.
>From that, we can get the cross section at the tree-level with
both polarized and unpolarized gluons.

\par
\begin{flushleft} {\bf III. SUSY QCD corrections (non-SM) to the
              subprocess $gg\rightarrow t\bar{t}$} \end{flushleft}
\par
1. Relevant Lagrangian in the MSSM.
\par
  The difference between the MSSM QCD and the SM QCD corrections stems from
the interactions of SUSY particles. Thus we can divide SUSY QCD corrections
into a standard and a non-standard part. The Lagrangian density of the
non-SM part of the SUSY QCD interaction is written as:
$$
\begin{array} {lll}
 L&=L_{1}+L_{2}+L_{3}+L_{4}
\end{array}
\eqno {(3.a.1)}
$$
Where
$$
\begin{array} {lll}
 L_{1}&=-i g_{s} A^{\mu}_{a} T_{jk}^{a} (\tilde{q}^{j}_{L}
          \partial_{\mu} \tilde{q}^{k}_{L}-
          \tilde{q}^{k}_{L} \partial_{\mu} \tilde{q}^{j}_{L})+(L \rightarrow R)
\end{array}
\eqno {(3.a.2)}
$$
$$
\begin{array} {lll}
 L_{2}&=-\sqrt{2} \hat{g_{s}} T_{jk}^{a} (\bar{\tilde{g}}_{a} P_{L} q^{k}
        \tilde{q}^{j\ast}_{L} +
        \bar{q}^{j} P_{R} \tilde{g}_{a} \tilde{q}^{k}_{L} -
        \bar{\tilde{g}}_{a} P_{R} q^{k} \tilde{q}_{R}^{j\ast} -
        \bar{q}^{j} P_{L} \tilde{g}_{a} \tilde{q}^{k}_{R})
\end{array}
\eqno {(3.a.3)}
$$
$$
\begin{array} {lll}
L_{3}&=\frac{i}{2} g_{s} f_{abc} \bar{\tilde{g^{a}}} \gamma_{\mu} \tilde{g^{b}} A_{\mu}^{c}
\end{array}
\eqno {(3.a.4)}
$$
$$
\begin{array} {lll}
L_{4}&=\frac{1}{6} g_{s}^{2} A_{\mu}^{a} A^{\mu}_{a} (\tilde{q}_{L}^{\ast}
      \tilde{q}_{L}+ \tilde{q}_{R}^{\ast} \tilde{q}_{R})+\frac{1}{2} g_{s}^2
      d_{abc} A_{\mu}^{a} A^{\mu b} (\tilde{q}_{L}^{i\ast} T_{ij}^{c}
      \tilde{q}^{j}_{L}+\tilde{q}_{R}^{i\ast} T_{ij}^{c} \tilde{q}^{j}_{R})
\end{array}
\eqno {(3.a.5)}
$$
q stands for quark, $\tilde{q}$ for corresponding squark, $\tilde{g}$ for
gluino, $P_{L}$ and $P_{R}$ for left, right helicity projections, respectively.
The mixing between the left- and right-handed stop quarks $\tilde{t}_{L}$
and $\tilde{t}_{R}$ can be very large due to the large mass of the top
quark, and the lightest scalar top-quark mass eigenstate $\tilde{t}_1$
can be much lighter than the top-quark and all the scalar partners of
the light quarks. Therefore the left-right mixing for the SUSY partners
of the top quark plays an important role. Here we only considered the
SUSY QCD effect from stop-quark, because we assume that other scalar
SUSY quarks are much heavier than the stop-quark and hence decoupled.
Furthermore we introduce the phase angle $\phi_{A}$ in the stop mixing matrix.
Defining $\theta$ as mixing angle of stop-quark, we have
$$
\begin{array} {lll}
\tilde{t}_{L}&=e^{\frac{-i \phi_{A}}{2}} (\tilde{t}_{1}\cos{\theta} +
\tilde{t}_{2} \sin{\theta} )
\end{array}
\eqno {(3.a.6)}
$$
$$
\begin{array} {lll}
\tilde{t}_{R}&=e^{\frac{i \phi_{A}}{2}} (-\tilde{t}_{1} \sin{\theta} +
\tilde{t}_{2} \cos{\theta} )
\end{array}
\eqno {(3.a.7)}
$$
where we suppose
$m_{\tilde{t}_{1}} \leq m_{\tilde{t}_{2}}$.

\par
2. Analytical results of the MSSM QCD corrections.
\par
The one-loop SUSY QCD correction diagrams are shown in Fig.1(b). In the
following we present only the amplitude expressions of s-channel and
t-channel. The amplitude of u-channel can be obtained from the t-channel
expression by the following variable exchanges: $t \leftrightarrow u$,
$k_1 \leftrightarrow k_2$, $\epsilon_{\mu}^{a} (k_1) \leftrightarrow
\epsilon_{\nu}^{b} (k_2)$ and $T^{a} \leftrightarrow T^{b}$.
The one-loop diagrams can be divided into three groups: the self-energy
diagrams of gluon and top-quark shown in Fig.1(b.1); $gtt$ and $ggg$ vertex
correction diagrams shown in Fig.1(b.2); box diagrams shown in Fig.1 (b.3).
The ultraviolet divergence is controlled by dimensional regularization
($n=4-\epsilon$). The strong coupling-constants are renormalized by using
the modified Minimal Subtraction ($\bar{MS}$) scheme at
charge-renormalization scale $\mu_{R}$. This scheme violates SUSY explicitly
and the $q\tilde{q}\tilde{g}$ Yukawa coupling $\hat{g}_s$, which should be
the same with the $qqg$ gauge coupling $g_s$ in supersymmetry, takes a finite
shift at one-loop order. Therefore we take this shift between $\hat{g}_s$
and $g_s$ as shown in Eq.(3.b.1) into account in our calculation, in order
to have the
physical amplitudes independent of the renormalization scheme and
we subtract the contribution of the false, non-supersymmetric degrees of
freedom (also called $\epsilon$ scalars) \cite{Martin}.
$$
\hat{g}_s=g_s \left [ 1+ \frac{\alpha_s}{4 \pi}
(\frac{2}{3} C_A-\frac{1}{2} C_F) \right] ,
\eqno {(3.b.1)}
$$
 where $C_A=3$ and $C_F=4/3$ are the Casimir invariants of SU(3) gauge group.
The heavy particles(top quarks, gluino, stop-quarks, etc.) are removed from
the $\mu_{R}$ evolution of $\alpha_{s}(\mu_{R}^{2})$, then they are decoupled
smoothly when momenta are smaller than their masses\cite{s8}. We define
masses of heavy particles as pole masses.

\par
The renormalized amplitude corresponding to all SUSY QCD
one-loop corrections (as shown in Fig.1) can be split into the following
components:
$$
\begin{array} {lll}
\delta M = \delta M_{s}+ \delta M_{v} + \delta M_{box} + \delta M_{d}.
\end{array}
\eqno {(3.b.2)}
$$
where $\delta M_{s}$, $\delta M_{v}$, $\delta M_{box}$ and $\delta M_{d}$
are the one-loop amplitudes corresponding to the self-energy, vertex,
box correction diagrams and the decoupling part, respectively.
The $\delta M_{d}$ stems from the decoupling of the heavy flavors from the
running strong coupling, and is given explicately by (see also
\cite{Martin} \cite{s8}):
$$
\begin{array} {lll}
\delta M_{d} ~ =
        M_{0} (\frac{\alpha_{s}(\mu)}{\pi})
        [\frac{1}{24} \log (\frac{\mu_{R}^{2}}{m_{\tilde{t}_{1}}^2})
        + \frac{1}{24} \log (\frac{\mu_{R}^{2}}{m_{\tilde{t}_{2}}^2})+
        + \frac{1}{6} \log (\frac{\mu_{R}^{2}}{m_{t}^2})+
         \frac{1}{2} \log (\frac{\mu_{R}^{2}}{m_{\tilde{g}}^2})]
\end{array}
\eqno {(3.b.3)}
$$
\par
3.Self-energy corrections to the amplitude.
\par
The amplitude of self-energy diagrams $\delta M_{s}$
(Fig.1.(b.1)) can be decomposed into $\delta M_{s}^{g}$ (gluon self-energy)
and $\delta M_{s}^{q}$ (top-quark self-energy), i.e.
$$
\begin{array} {lll}
\delta M_{s} &=~ \delta M_{s}^{g} + \delta M_{s}^{q} \\
             &=~ \delta M_{s}^{g(s)} + \delta M_{s}^{g(t)} +
       & \delta M_{s}^{g(u)}+ \delta M_{s}^{q(t)} + \delta M_{s}^{q(u)}.
\end{array}
\eqno {(3.c.1)}
$$
The amplitudes $\delta M_{s}^{g(s)}$, $\delta M_{s}^{g(t)}$ and
$\delta M_{s}^{g(u)}$ are for s-, t- and u-channel, respectively. They can
be expressed as:
$$
\begin{array} {lll}
\delta M_{s}^{g(s)} = \frac{1}{2} M_{0}^{(s)}
                         [\Pi(k_{1}^{2})+\Pi(k_{2}^{2})+2 \Pi(s)],
\end{array}
\eqno {(3.c.2)}
$$
$$
\begin{array} {lll}
\delta M_{s}^{g(t)} = \frac{1}{2} M_{0}^{(t)}
                        [\Pi(k_{1}^{2})+\Pi(k_{2}^{2})],
\end{array}
\eqno {(3.c.3)}
$$
$$
\begin{array} {lll}
\delta M_{s}^{g(u)} = \frac{1}{2} M_{0}^{(u)}
                        [\Pi(k_{1}^{2})+\Pi(k_{2}^{2})] ,
\end{array}
\eqno {(3.c.4)}
$$
where $M_{0}$ is the tree-level amplitude defined in Eq (2.4).
$$
\begin{array} {lll}
\Pi(k^{2}) ~ = &~ -\frac{\alpha_{s}}{4 \pi}
               (T_{F}(\bar{B}_{0}+4 \bar{B}_{1}+4 \bar{B}_{21})[k,m_{\tilde{t}_{1}},m_{\tilde{t}_{1}}]+\\
             &  T_{F}(\bar{B}_{0}+4 \bar{B}_{1}+4 \bar{B}_{21})[k,m_{\tilde{t}_{2}},m_{\tilde{t}_{2}}]-
                4 C_{A} (\bar{B}_{1}+\bar{B}_{21})[k,m_{\tilde{g}},m_{\tilde{g}}]-\frac{1}{3} C_{A}).
\end{array}
\eqno {(3.c.5)}
$$
where $C_{F}=\frac{4}{3}$, $T_{F}=\frac{1}{2}$ , $C_{A}=3$ are invariants
in the SU(3) color group, $B_{i}$ and $B_{ij}$ are Passarino-Veltman
two-point functions
\cite{s19}\cite{s20}. The definitions of $\bar{B}_{0}$, $\bar{B}_{1}$ and
$\bar{B}_{21}$ are listed in Appendix A. The amplitude $\delta M_{s}^{q(t)}$
is written as:
$$
\begin{array} {lll}
\delta M_{s}^{q(t)} & = & \frac{-i g_{s}^2 T_{ik}^{a} T_{lj}^{b}}
   {(t-m_{t}^{2})^{2}}  \epsilon ^{\mu,a} (k_1) \epsilon ^{\nu,b} (k_2)
   \bar{u}_{i}(p_1) \gamma_{\mu}  \\
&& (\rlap/k_{2}-\rlap/p_{2}+m_{t}) \left[ \hat{\Sigma}_{kl} (k_{2}-p_{2})
   \right] (\rlap/k_{2}-\rlap/p_{2}+m_{t}) \gamma_{\nu}  v_{j}(p_2).
\end{array}
\eqno {(3.c.6)}
$$
Here we define
$$
\begin{array} {lll}
\hat{\Sigma}_{kl} (p) ~=~ C_{F}  (H_{L}  \rlap/p P_{L} +
          H_{R}  \rlap/p P_{R}- H^{S}_{L} P_{L} - H^{S}_{R} P_{R})
\delta_{kl}
\end{array}
\eqno {(3.c.7)}
$$
with
$$
\begin{array} {lll}
H_{L} ~=~ \frac{\hat{g}_{s}^{2}}{8 \pi ^{2}} x_{1} x_{3}
B_{1}[p,m_{\tilde{g}},m_{\tilde{t}_{1}}]+
        (m_{\tilde{t}_1} \rightarrow m_{\tilde{t}_{2}}, x_i \rightarrow y_i)+
        \frac{1}{2} (\delta Z_{L} +\delta Z_{L}^{\dag}),
\end{array}
\eqno {(3.c.8)}
$$
$$
\begin{array} {lll}
H_{R} ~=~\frac{\hat{g}_{s}^{2}}{8 \pi ^{2}} x_{2} x_{4}
B_{1}[p,m_{\tilde{g}},m_{\tilde{t}_{1}}]+
        (m_{\tilde{t}_1} \rightarrow m_{\tilde{t}_{2}}, x_i \rightarrow y_i)+
         \frac{1}{2} (\delta Z_{R} +\delta Z_{R}^{\dag}) ,
\end{array}
\eqno {(3.c.9)}
$$
$$
\begin{array} {lll}
H^{S}_{L} ~=~\frac{\hat{g}_{s}^{2}}{8 \pi ^{2}} x_{2} x_{3} m_{\tilde{g}}
B_{0}[p,m_{\tilde{g}},m_{\tilde{t}_{1}}]+
        (m_{\tilde{t}_1} \rightarrow m_{\tilde{t}_{2}}, x_i \rightarrow y_i)+
         \frac{1}{2} m_{t} (\delta Z_{L} +\delta Z_{R}^{\dag})+\delta m_{t},
\end{array}
\eqno {(3.c.10)}
$$
$$
\begin{array} {lll}
H^{S}_{R} ~=~\frac{\hat{g}_{s}^{2}}{8 \pi ^{2}} x_{1} x_{4} m_{\tilde{g}}
B_{0}[p,m_{\tilde{g}},m_{\tilde{t}_{1}}]+
        (m_{\tilde{t}_1} \rightarrow m_{\tilde{t}_{2}}, x_i \rightarrow y_i)+
          \frac{1}{2} m_{t} (\delta Z_{R} +\delta Z_{L}^{\dag})+\delta m_{t}.
\end{array}
\eqno {(3.c.11)}
$$
Where we abbreviate $\phi=\phi_{A}$,
 $x_{1}=\cos \theta e^{-i \phi}$,
 $x_{2}=\sin \theta e^{i \phi}$, $x_{3}=\cos \theta e^{i \phi}$,
 $x_{4}=\sin \theta e^{-i \phi}$,
 $y_{1}=\sin \theta e^{-i \phi}$,
 $y_{2}=- \cos \theta e^{i \phi}$, $y_{3}=\sin \theta e^{i \phi}$,
 $y_{4}=- \cos \theta e^{-i \phi}$, and $\theta$ is mixing angle
 of stop-quarks, see Eq (3.a.6 $\sim$ 7).

The explicit expressions of the top-quark wave-function renormalization
constants have the following forms:
$$
\begin{array} {lll}
\delta Z_{L} &=&
   - \frac{\hat{g}_{s}^{2}}{8 \pi ^{2}} (x_{1} x_{3} Re[B_{1}]- \frac
{m_{\tilde{g}}}{m_{t}} (x_{1} x_{4}-x_{2} x_{3}) Re[B_{0}]\\
   &+& m_{t}^{2} (x_{1} x_{3}+x_{2} x_{4}) Re[B_{1}^{'}] \\
&-& m_{t} m_{\tilde{g}} (x_{2} x_{3}+x_{1} x_{4}) Re[B_{0}^{'}])
         [p,m_{\tilde{g}},m_{\tilde{t}_{1}}]|_{p^{2}=m_{t}^{2}},
\end{array}
\eqno {(3.c.12)}
$$
$$
\begin{array} {lll}
\delta Z_{R} &=&
         - \frac{\hat{g}_{s}^{2}}{8 \pi ^{2}} (x_{2} x_{4} Re[B_{1}]+
m_{t}^{2} (x_{1} x_{3}+x_{2} x_{4})
         Re[B_{1}^{'}]  \\
     &-& m_{t} m_{\tilde{g}} (x_{2} x_{3}+x_{1} x_{4}) Re[B_{0}^{'}])
         [p,m_{\tilde{g}},m_{\tilde{t}_{1}}]|_{p^{2}=m_{t}^{2}},
\end{array}
\eqno {(3.c.13)}
$$
$$
\begin{array} {lll}
\delta m_{t} &=&
         \frac{\hat{g}_{s}^{2}}{16 \pi ^{2}}  ((x_{1} x_{3}+x_{2} x_{4})
m_{t} Re[B_{1}]\\
         &-&(x_{2} x_{3}+x_{1} x_{4}) m_{\tilde{g}} Re[B_{0}])
         [p,m_{\tilde{g}},m_{\tilde{t}_{1}}]|_{p^{2}=m_{t}^{2}},
\end{array}
\eqno {(3.c.14)}
$$
We use the following abbreviations:
 $B_{i,ij}^{'}[p,m_{1},m_{2}]=\frac{\partial B_{i,ij}[p,m_{1},m_{2}]}
 {\partial p^2}$.

\par
4.Vertex-corrections to the amplitude.
\par
The amplitudes for vertex diagrams can be expressed as:
$$
\begin{array} {lll}
\delta M_{v}^{(l)} ~ = ~ g_{s} \epsilon ^{\mu,a} (k_1) \epsilon ^{\nu,b} (k_2)
    \bar{u}_{i}(p_1) \Lambda^{(l)}v_{j}(p_2),~~~(l=s,t,u), \\
\end{array}
\eqno {(3.d.1)}
$$
where
$$
\begin{array} {lll}
\Lambda^{(s)} ~=& ~ -\frac{T^{c}_{ij}}{s}
  \left[ \Lambda_{\mu\nu\rho}^{(3g)}(k_1,k_2) \right] \gamma_{\rho}\\
 &-\frac{f_{abc}}{s} [(k_{1} - k_{2})_{\rho} g_{\mu\nu}+
  (2 k_{2} + k_{1})_{\mu} g_{\nu\rho}  \\
 &-(2 k_{1} + k_{2})_{\nu} g_{\mu\rho}]
  \left[ \Lambda_{\rho,(ij)}^{c}(p_{1},p_{2}) \right],
\end{array}
\eqno {(3.d.2)}
$$
and
$$
\begin{array} {lll}
  \Lambda^{(t)} ~=& ~ \frac{-i}{t-m_t^2}
  \left\{ T^{b}_{mj} \left[ \Lambda_{\mu,(im)}^{a} (p_{1},k_{1}-p_{1})\right]
  (\rlap/k_{2}-\rlap/p_{2}+m_{t})\gamma _{\nu } \right. \\
  & \left. + T^{a}_{im} \gamma _{\mu } (\rlap/k_{2}-\rlap/p_{2}+m_{t})
  \left[ \Lambda_{\nu,(mj)}^{b}(k_2-p_2,p_2) \right] \right\}.
\end{array}
\eqno {(3.d.3)}
$$
The functions $\Lambda_{\mu\nu\rho}^{(3g)}$ and $\Lambda_{\mu,(ij)}^{a}$
are listed in Appendix B.
\par
5. Box-corrections to the amplitude.
\par
The box diagram corrections in the t-channel (Fig.1(b.3)) are given as
follows:
$$
\begin{array} {lll}
\delta M_{box}^{(t)} ~ =& ~2 g_{s}^2 \epsilon ^{\mu,a} (k_1) \epsilon ^{\nu,b} (k_2) \bar{u}_{i}(p_1)
   ((T^{c} T^{a} T^{b} T^{c})_{ij} F_{\mu\nu}^{(t1)}\\
 &- i f_{bcd} (T^{c} T^{a} T^{d})_{ij} F_{\mu\nu}^{(t2)}-
   f_{acm} f_{bmd} (T^{c} T^{d})_{ij} F_{\mu\nu}^{(t3)} \\
 &- [T^{c} (T^{a} T^{b}+T^{b} T^{a}) T^{c}]_{ij} F_{\mu\nu}^{(t4)})
   v_{j}(p_2),
\end{array}
\eqno {(3.e.1)}
$$
Where $f_{abc}$ is defined as $[T^{a},T^{b}]=if_{abc}T^{c}$.
The form factors $F_{\mu\nu}^{(ti)}(i=1 - 4)$ correspond to the kernel
of the four Feynman diagrams in Fig.1(b.3) respectively and are given
explicitly in Appendix C.
\par
\vskip 20mm
6. Total cross section.
\par
Collecting all terms in Eq (3.b.2), we can get the total cross section:
$$
\begin{array} {lll}
\sigma (\lambda_1, \lambda_2) &= \sigma_{0}(\lambda_1, \lambda_2)
(1+\delta \sigma (\lambda_1, \lambda_2))\\
&=\frac{1}{16 \pi s^2 }
             \int_{t^{-}}^{t^{+}} dt {\sum_{spins}^{}}
             [|M_{0}|^{2}+2 Re(M_{0}^{\dagger} \delta M)]
\end{array}
\eqno {(3.f.1)}
$$
where $t^\pm=(m_t^2-\frac{1}{2}s)\pm\frac{1}{2}s \beta_t$,
$\beta_t=\sqrt{1-4m_t^2/s}$, and the spin sum is performed only over the
final top-quark pair when we considered polarized gluons.

\begin{flushleft} {\bf IV. Numerical results} \end{flushleft}
\par
   We denote $\hat{\sigma}_{0}$ for the Born cross section and
$\hat{\sigma}$ for the cross section including one-loop SUSY QCD corrections
of subprocess $gg \rightarrow t \bar{t}$, and define its relative correction
as $\hat{\delta} = \frac{\hat{\sigma} - \hat{\sigma}_{0}}{\hat{\sigma}_{0}}$.
For polarized gluon fusions, $\hat{\sigma}_{++}$, $\hat{\sigma}_{--}$ and
$\hat{\sigma}_{+-}$ are the cross sections with positive, negative and mixed
polarization of the gluons, respectively. In order to inspect the CP violating
effects we introduce the CP-violation parameter for the subprocess defined by
$\hat{\xi}_{CP}=\frac{\hat{\sigma}_{++} - \hat{\sigma}_{--}}
{\hat{\sigma}_{++} + \hat{\sigma}_{--}}$. The possible SUSY QCD effects in
$gg \rightarrow t\bar{t}$ should be observed in $pp$ colliders. By analogy
we can define also the relative correction and CP violating parameter for
the process $pp \rightarrow gg \rightarrow t \bar{t}$ as $\delta =
\frac{\sigma - \sigma_{0}}{\sigma_{0}}$ and $\xi_{CP}=\frac{\sigma_{++} -
\sigma_{--}}{\sigma_{++} + \sigma_{--}}$, respectively. The SUSY QCD
contribution to the process $p(P_{1},x)p(P_{2},y) \rightarrow gg \rightarrow
t \bar {t} X$ (x,y are polarizations of protons) can be obtained by convoluting
the subprocess with gluon distribution functions.
$$
\begin{array} {lll}
\sigma (s) &= \int dx_{1} dx_{2} G(x_{1},Q) G(x_{2},Q)
              \hat{\sigma}(\hat{s},\alpha_{s}(\mu))
\end{array}
\eqno {(4.1)}
$$
with $k_{1}=x_{1}P_{1}$, $k_{2}=x_{2}P_{2}$
and $\tau=x_{1}x_{2}=\hat{s}/s$.
$G(x_{i},Q)(i=1,2)$ are gluon distribution functions of protons.
We take $Q=\mu_{R}=2m_{t}$.
\par
In order to get results of top quark pair production from
polarized $pp$ collisions, we need to consider the polarized
gluon distributions in protons. The cross sections of polarized
$pp \rightarrow gg \rightarrow t \bar{t} X$ can be written as
$$
\begin{array} {lll}
\sigma (x,y) &= \Sigma_{\lambda_{1},\lambda_{2} =\pm}
                \int dx_{1} dx_{2} G^{x\lambda_{1}}(x_{1},Q) G^{y\lambda_{2}}(x_{2},Q)
              \hat{\sigma}_{\lambda_{1},\lambda_{2}}(\hat{s},\alpha_{s}(\mu))
\end{array}
\eqno {(4.2)}
$$
where x and y are the polarizations of incoming protons and
$\lambda_{1},~~ \lambda_{2}$ are the polarizations of gluons
inside protons. $G^{x\lambda_{1}}(x,Q),~G^{y\lambda_{2}}(x,Q)=G^{\pm}(x,Q)$
for equal (+) and opposite (-) polarization,
$G^{+}(x,Q)$ and $G^{-}(x,Q)$ are polarized gluon distribution functions in
the proton.
\par
We used unpolarized proton structure functions of Gl\"uck et al.
\cite{Glk} in our numerical calculations. For the polarized proton
structure functions, we use the evolution equations of Gl\"uck et al.
\cite{GLK} with input parameters from the paper of Stratmann et al.
\cite{STR} (Next-To-Leading-Order). Since the structure functions are
one of the least certain input in our calculation, we checked the result
against other set, i.e. the polarized structure functions $G^{\pm}(x,Q)$
of Brodsky et. al \cite{Geh} (Using Leading-Order only). This tests the
stability of our results against the
particular form of the input structure functions. The two different
sets of input are compared in Fig.2, which gives the relative SUSY QCD
 correction ($\delta$) and $\xi_{CP}$ versus c.m. energy $\sqrt{s}$ for
the process $pp \rightarrow gg \rightarrow t \bar{t} X$. Though the SUSY
QCD corrections from the two sets of structure functions are not too
different, for $\delta$, there is some noticable change for $\xi_{CP}$.
Because $\xi_{CP}$ depends strongly on the c.m. energy of the subprocess
$gg \rightarrow t\bar{t}$ (shown in Fig.3(b)), a small
modification of structure functions may lead to a large change of
$\xi_{CP}$. Thus we can infer that the NLO-QCD calculation is required 
and the precise numerical prediction does depend on the reliability of
the structure functions.   

\par
The SUSY QCD relative corrections are about $2 \% \sim 4 \%$ and decrease
with increasing c.m. energy from Fig.2. These correction effects are
within reach
of future precision experiments and provide a possible discrimination of
the SM and the MSSM effects. From Fig.2(c) we can see that the CP
violation parameter $\xi_{CP}$ can be $10^{-3}$.
Therefore, CP violation in this process stemming from the SUSY QCD can
in principle be tested in future precision experiments. That would help
us to learn more about the sources of CP violation.
\par
In order to explore the effects of the SUSY QCD correction for future
arrangements of optimal experimental conditions, we also investigate
the subprocess $gg \rightarrow t \bar{t}$.
\par
The relative SUSY QCD correction and CP violating parameter versus c.m. energy
($\sqrt{\hat{s}}$) for different polarization gluons are plotted in Fig.3
($a \sim c$) with $m_{\tilde{g}}~=~200~GeV $, $m_{\tilde{t}_1}~=~ 250~GeV $,
$m_{\tilde{t}_2}~ =~ 450 ~GeV $, and $\theta~ =~ \phi~ =~ 45^{\circ}$.
In Fig.3(a) $\hat{\delta}_{++}$ and $\hat{\delta}_{--}$ are
drawn in solid line and dashed line, respectively. $\hat{\xi}_{CP}$ as a
function of c.m. energy is depicted in Fig.3(b) and $\hat{\delta}_{+-}$
as function of $\sqrt{\hat{s}}$ is plotted in Fig.3(c). Each curve in
Fig.3(a) has an obvious peak near the position of the threshold of
top pair production. That large enhancement is the combined effect of
the threshold, when $\sqrt{\hat{s}}$ is just larger than
$2 m_t = 350~GeV$, and the resonance when $\sqrt{\hat{s}}
\sim 2 m_{\tilde{g}} = 400~GeV$. The small spikes around the position of
$\sqrt{\hat{s}}=900~GeV$, there $\sqrt{\hat{s}} \sim 2 m_{\tilde{t}_2} =
900~GeV$, shows also the resonance effect. Although Fig.3(a) shows that
$\hat{\delta}_{++}$ and $\hat{\delta}_{--}$ approach equal
values when the c.m. energy is far beyond its threshold value $2 m_{t}$,
the quantitative difference between $\hat{\delta}_{+-}$
and $\hat{\delta}_{++}$ still exists in the whole energy range plotted in
these figures. Fig.3(b) shows also that $\hat{\xi}_{CP}$ will be zero,
if the c.m. energy is below the threshold of SUSY particles in the loop(
i.e. $\sqrt{\hat{s}} \le 2 m_{\tilde{g}}~=~400~GeV $ in Fig.3(b)).
This is reasonable because only beyond this point can we
have absorptive terms which give contributions to $\hat{\xi}_{CP}$.
$\hat{\xi}_{CP}$ has obvious resonance effect in the regions around
$\sqrt{\hat{s}} \sim 2 m_{\tilde{g}}~=~400 GeV$ and $\sqrt{\hat{s}} \sim
2 m_{\tilde{t}_i}, (i=1,2) =500~GeV,~ 900~GeV$. We also find that the two
stop quarks give opposite contributions to $\hat{\xi}_{CP}$
and when their masses
are degenerate $\hat{\xi}_{CP}$ will vanish. When the c.m. energy
$\sqrt{\hat{s}}$ is larger than $1~TeV$, $\hat{\xi}_{CP}$ will be
near zero, because the contributions from the two stop quarks will cancel
each other. Therefore a quantitative strong change of $\hat{\xi}_{CP}$
as function of c.m. energy can be an indication for the
signals of stop quarks and gluino.
\par
$\hat{\sigma}(\pm,\pm)$ and $\hat{\xi}_{CP}$ as functions of $m_{\tilde{g}}$
are shown in Fig.4 (a) and Fig.4 (b), respectively. In Fig.4 we take
$\sqrt{\hat{s}} = 500 GeV$, $m_{\tilde{t}_1} = 100 GeV $, $m_{\tilde{t}_2}
= 450 GeV $, $\theta = \phi =45^{\circ}$. We can see from Fig.4(b) that
$\hat{\xi}_{CP}$ changes its sign when $m_{\tilde{g}}$ is near $m_{t}=175~GeV$.
The curves in Fig.4(a)(b) show again the resonance effect when
$\sqrt{\hat{s}} \sim 2 m_{\tilde{g}} = 500~GeV$, note that for
each
line there is a steep change of the value of $\hat{\sigma}(\pm,\pm)$
or $\hat{\xi}_{CP}$ around the position of $m_{\tilde{g}}=250 GeV$.
\par
Dependences of relative correction $\hat{\delta}_{\pm \pm}$ and $\hat{\xi}_{CP}$
for the subprocess $gg \rightarrow t \bar{t}$
on $m_{\tilde{t}_{1}}$ are plotted in Fig.5 (a) and Fig.5 (b).
$\hat{\delta}_{\pm \pm}$ and $\hat{\xi}_{CP}$ as functions of
$m_{\tilde{t}_{2}}$ are shown in Fig.6 (a) and Fig.6 (b),
respectively.
In all figures of Fig.5 and Fig.6, we take the common parameter set with
$\sqrt{\hat{s}} = 500~GeV$, $m_{\tilde{g}}=200~GeV$ and $\theta = \phi
=45^{\circ}$. In Fig.5, we set $m_{\tilde{t}_{2}}=450~GeV$, whereas
$m_{\tilde{t}_{1}}=100~GeV$ in Fig. 6.
We find that $\hat{\xi}_{CP}$ in fact increases with mass splitting of
stop-quarks (i.e. $m_{\tilde{t}_{2}}-m_{\tilde{t}_{1}}$) and
when $m_{\tilde{t}_{1}}=m_{\tilde{t}_{2}}$, $\hat{\xi}_{CP}$ is equal to zero.
The resonance effect of stop quarks, when $\sqrt{\hat{s}} \sim 2~m_{\tilde{t}_{i}}
(i=1,2)$, is superimposed on the curves in Fig.5 (a)(b) and Fig. 6(a)(b)
around the positions of $m_{\tilde{t}_{1}}=250~GeV$ in Fig.5(a)(b) and
$m_{\tilde{t}_{2}}=250~GeV$ in Fig.6(a)(b), respectively.
Around those points the relatively sharp changes of the values
of $\hat{\xi}_{CP}$ and the relative corrections are shown in these figures.
\par
 Finally, the dependence of $\hat{\delta}_{\pm \pm}$ and
$\hat{\xi}_{CP}$ on the phase $\phi$ is shown in Fig.7 (a) and (b).
In Fig.7, we take $\sqrt{\hat{s}} = 500~GeV$,
$m_{\tilde{g}}=200GeV$, $\theta =45^{\circ}$ and $m_{\tilde{t}_{1}}=150~GeV$.
We find that $\hat{\xi}_{CP}$ is directly proportional to $\sin{(2\phi)}$ and
reaches its maximal value when $\phi=\frac{\pi}{4}$.

\vskip 5mm
\begin{flushleft} {\bf IV. Conclusion} \end{flushleft}
In this work we have studied the one-loop supersymmetric QCD corrections to
the subprocess $gg \rightarrow t\bar{t}$ and process
$pp \rightarrow gg \rightarrow t \bar{t}X$.
The calculations show that the SUSY QCD effects are significant.
The absolute values of the corrections are about
$2\%\sim 4 \%$, so they may be observable in future precision experiments.
Furthermore, we find $\xi_{CP}$ depends strongly on masses of SUSY particles
and can reach $10^{-3}$ when we take plausible SUSY parameters.
\par
The results show that there is an obvious difference between the corrections
for the protons polarized with parallel spin and that with
anti-parallel spin. Hence there is a possibility to study spin-dependence
in the frame of the MSSM QCD.
\par
We also presented and discussed the results of the subprocess $gg \rightarrow
t \bar{t}$. We find that when the c.m. energy passes through the value
$2 m_{\tilde{g}}$ or $2 m_{\tilde{t}_i}$ ($i=1,2$), the value of the CP
violating parameter $\hat{\xi}_{CP}$ changes strongly. If c.m. energy is less
than both $2 m_{\tilde{g}}$ and $2 m_{\tilde{t}_{i}}~(i=1,2)$,
$\hat{\xi}_{CP}$ will be zero. If in future experiments a sharp change in
$\hat{\xi}_{CP}$
is found with $\sqrt{\hat{s}}$ running from low c.m. energy to
high c.m. energy, it would be interpreted as a
signal of SUSY particles. Furthermore, because the CP violating parameter
$\hat{\xi}_{CP}$ is sensitive on the mass of gluino(as shown in Fig.4 (b))
and the mass splitting of stop-quarks $m_{\tilde{t}_{2}}-m_{\tilde{t}_{1}}$
(as shown in Fig.5 and Fig.6), we can also get information of SUSY particles
from precise measurements of $\hat{\xi}_{CP}$.
\par
The authors would like to thank Prof. A. Bartl for useful discussions and
comments. One of the authors, Yu Zeng-Hui, would like to thank Prof.
H. Stremnitzer for his help.
\newpage
\begin{center} Appendix\end{center}

\par
A. Loop integrals:
\par
We adopt the definitions of two-, three- and four-point one-loop
Passarino-Veltman integral functions of reference\cite{s19}\cite{s20}.
\par
1.The two-point integrals are:
$$
\{B_0;B_{\mu};B_{\mu\nu}\}(p,m_1,m_2)=
{\frac{(2\pi\mu)^{4-n}}{i\pi^2}}\int d^n q
{\frac{\{1;q_{\mu};q_{\mu}q_{\nu}\}}{[q^2-m_1^2][(q+p)^2-m_2^2]}},
~~~~~(A.a.1)
$$
The function $B_{\mu}$ is proportional to $p_{\mu}$:
$$
B_{\mu}(p,m_{1},m_2)=p_{\mu} B_{1}(p,m_1,m_2)
~~~~~(A.a.2)
$$
Similarly we define:
$$
B_{\mu\nu}=p_{\mu}p_{\nu} B_{21}+g_{\mu\nu} B_{22}
~~~~~(A.a.3)
$$
We denote $\bar{B}_{0}= B_{0}-\Delta$, $\bar{B}_{1}= B_{1}+\frac{1}{2}\Delta$
and $\bar{B}_{21}= B_{21}-\frac{1}{3}\Delta$. with $\Delta= \frac{2}{\epsilon}
-\gamma +\log (4\pi)$, $\epsilon=4-n$. ${\mu}$ is the scale parameter.
\par
2. The three-point integrals are:
$$
\{C_0;C_{\mu};C_{\mu\nu};C_{\mu\nu\rho}\}(p,k,m_1,m_2,m_3)=
$$
$$
-{\frac{(2\pi\mu)^{4-n}}{i\pi^2}}\int d^n q
{\frac{\{1;q_{\mu};q_{\mu}q_{\nu};q_{\mu}q_{\nu}q_{\rho}\}}
{[q^2-m_1^2][(q+p)^2-m_2^2][(q+p+k)^2-m_3^2]}},
~~~~~(A.a.4)
$$
We define form-factors as follows:
$$
C_{\mu}=p_{\mu} C_{11} + k_{\mu} C_{12}
$$
$$
C_{\mu\nu}=p_{\mu} p_{\nu} C_{21}+k_{\mu}k_{\nu} C_{22}+
(p_{\mu}k_{\nu}+k_{\mu}p_{\mu}) C_{23}+ g_{\mu\nu} C_{24}
$$
$$
C_{\mu\nu\rho}=p_{\mu}p_{\nu}p_{\rho} C_{31}+
               k_{\mu}k_{\nu}k_{\rho} C_{32}+
               (k_{\mu}p_{\nu}p_{\rho} +
                p_{\mu}k_{\nu}p_{\rho} +
                p_{\mu}p_{\nu}k_{\rho}) C_{33}+
$$
$$
               (k_{\mu}k_{\nu}p_{\rho} +
                p_{\mu}k_{\nu}k_{\rho} +
                k_{\mu}p_{\nu}k_{\rho}) C_{34}+
                (p_{\mu} g_{\nu\rho}+p_{\nu} g_{\mu\rho}+
               p_{\rho} g_{\mu\nu}) C35+
$$
$$
                (k_{\mu} g_{\nu\rho}+k_{\nu} g_{\mu\rho}+
                k_{\rho} g_{\mu\nu}) C36
~~~~~~~~~~~~~~~(A.a.5)
$$

\par
3. The four-point integrals are:
$$
\{D_0;D_{\mu};D_{\mu\nu};D_{\mu\nu\rho};D_{\mu\nu\rho\alpha}\}
(p,k,l,m_1,m_2,m_3,m_4)=
$$
$$
{\frac{(2\pi\mu)^{4-n}}{i\pi^2}}\int d^n q
{\frac{\{1;q_{\mu};q_{\mu}q_{\nu};q_{\mu}q_{\nu}q_{\rho};q_{\mu}q_{\nu}q_{\rho}q_{\alpha}\}}
{[q^2-m_1^2][(q+p)^2-m_2^2][(q+p+k)^2-m_3^2][(q+p+k+l)^2-m_4^2]}},
~~~~~(A.a.6)
$$
Again we define form-factors of D functions:
$$
D_{\mu}=p_{\mu} D_{11}+k_{\mu} D_{12}+l_{\mu} D_{13}
$$
$$
D_{\mu\nu}=p_{\mu}p_{\nu} D_{21} +k_{\mu}k_{\nu} D_{22} +l_{\mu}l_{\nu}
D_{23}+\\
\{pk\}_{\mu\nu} D_{24}+ \{pl\}_{\mu\nu} D_{25}+ \{kl\}_{\mu\nu}
D_{26}+g_{\mu\nu}D_{27}
$$
$$
D_{\mu\nu\rho}=p_{\mu}p_{nu}p_{\rho}D_{31} +
               k_{\mu}k_{nu}k_{\rho}D_{32}+
               l_{\mu}l_{nu}l_{\rho}D_{33}+
               \{kpp\}_{\mu\nu\rho} D_{34}+
  $$
  $$
               \{lpp\}_{\mu\nu\rho} D_{35}+
               \{pkk\}_{\mu\nu\rho} D_{36}+
               \{pll\}_{\mu\nu\rho} D_{37}+
               \{lkk\}_{\mu\nu\rho} D_{38}+
$$
$$
               \{kll\}_{\mu\nu\rho} D_{39}+
               \{pkl\}_{\mu\nu\rho} D_{310}+
               \{pg\}_{\mu\nu\rho} D_{311}+
               \{kg\}_{\mu\nu\rho} D_{312}+
               \{lg\}_{\mu\nu\rho} D_{313} ~~~~~~(A.a.7)
$$
\par
where
$$
\{pk\}_{\mu\nu}=p_{\mu}k_{\nu}+k_{\mu}p_{\nu}
$$
$$
\{pkl\}_{\mu\nu\rho}=p_{\mu}k_{\nu}l_{\rho}+l_{\mu}p_{\nu}k_{\rho}+
k_{\mu}l_{\nu}p_{\rho}
$$
$$
\{pg\}_{\mu\nu\rho}=p_{\mu}g_{\nu\rho}+p_{\nu}g_{\mu\rho}+p_{\rho}g_{\mu\nu}
~~~~~(A.a.8)
$$
\par
The numerical calculation of the vector and tensor loop integral functions
can be traced back to the four scalar loop integrals $A_0$, $B_0$, $C_0$
and $D_0$ in Ref.\cite{s19}\cite{s20} and the references therein.
\par

B. Vertex corrections:
\par
The 3-gluon-vertex can be written as: (a,b,c are the color indices of the
external gluons)
$$
\begin{array} {lll}
\Lambda_{\mu\nu\rho}^{(3g)}(k_1,k_2)=
\frac{i g_{s}^3}{16 \pi^2} \left\{ Tr(T^{b} T^{c} T^{a})
\left[ \Lambda_{\mu\nu\rho}^{(1)}(k_1,k_2) \right] +
i f^{cmn} f^{anl} f^{blm} \left[ \Lambda_{\mu\nu\rho}^{(2)}(k_1,k_2)
\right] \right\},
\end{array}
\eqno {(A.b.1)}
$$
the vertex functions $\Lambda_{\mu\nu\rho}^{(1)},
\Lambda_{\mu\nu\rho}^{(2)}$ are expressed as follows:
$$
\begin{array} {lll}
\Lambda_{\mu\nu\rho}^{(a)}(k_1,k_2)=&  f_{1}^{(a)} g_{\mu\rho} k_{1 \nu}+
                   f_{2}^{(a)} g_{\mu\nu} k_{1 \rho}+
                   f_{3}^{(a)} g_{\nu\rho} k_{2 \mu}+
                   f_{4}^{(a)} g_{\mu\nu} k_{2 \rho}\\
                &+ f_{5}^{(a)} k_{1\nu} k_{1 \rho} k_{2 \mu}+
                   f_{6}^{(a)} k_{1\nu} k_{2 \rho} k_{2 \mu}\\
&+(m_{\tilde{t}_1} \rightarrow m_{\tilde{t}_2}, x_i \rightarrow y_i),
\end{array}
\eqno {(A.b.2)}
$$
where $a=1,2$, and the $f_{i}^{(1)},f_{i}^{(2)}$ are given in
terms of the Passarino-Veltman functions
with internal stop
lines $C_{ij}^{(1)}(=C_{ij}[-k_1,-k_2,m_{\tilde{t}_{1}},
m_{\tilde{t}_{1}},m_{\tilde{t}_{1}}])$ and internal gluino lines
$C_{ij}^{(2)}(=C_{ij}[-k_1,-k_2,m_{\tilde{g}},
m_{\tilde{g}},m_{\tilde{g}}])$. For simplicity, we abbreviate the definite
part of C integral functions (using the definitions of
\cite{s19}\cite{s20}) as follows:
$\bar{C}_{24}^{(a)}=C_{24}^{(a)}-\frac{1}{4}\Delta$,
$\bar{C}_{35}^{(a)}=C_{35}^{(a)}+\frac{1}{6} \Delta$,
$\bar{C}_{36}^{(a)}=C_{35}^{(a)}+\frac{1}{12} \Delta$. ($a=1,2$)

$$
\begin{array} {lll}
f_{1}^{(1)} = -8 \bar{C}_{24}^{(1)}-8 \bar{C}_{35}^{(1)},
\end{array}
$$
$$
\begin{array} {lll}
f_{2}^{(1)} = -4 \bar{C}_{24}^{(1)}-8 \bar{C}_{35}^{(1)},
\end{array}
$$
$$
\begin{array} {lll}
f_{3}^{(1)} = -8 \bar{C}_{36}^{(1)},
\end{array}
\eqno {(A.b.3)}
$$
$$
\begin{array} {lll}
f_{4}^{(1)} = -4 \bar{C}_{24}^{(1)}-8 \bar{C}_{36}^{(1)},
\end{array}
$$
$$
\begin{array} {lll}
f_{5}^{(1)} = 4 C_{12}^{(1)}+12 C_{23}^{(1)}+8 C_{33}^{(1)},
\end{array}
$$
$$
\begin{array} {lll}
f_{6}^{(1)} = 4 C_{12}^{(1)}+8 C_{22}^{(1)}+4 C_{23}^{(1)}+8 C_{34}^{(1)},
\end{array}
$$
and
$$
\begin{array} {lll}
f_{1}^{(2)} =&   -8 m_{\tilde{g}}^{2} C_{0}^{(2)}-4 m_{\tilde{g}}^{2} C_{11}^{(2)}-
                16 \bar{C}_{24}^{(2)}+12 \epsilon  C_{24}^{(2)}-
                 8 \bar{C}_{35}^{(2)}+6 \epsilon  C_{35}^{(2)}\\
            &   + 8 k_1 \cdot k_2 C_{12}^{(2)} +
                16 k_1 \cdot k_2 C_{23}^{(2)} +
                 8 k_1 \cdot k_2 C_{33}^{(2)},
\end{array}
$$
$$
\begin{array} {lll}
f_{2}^{(2)} =  -4 m_{\tilde{g}}^{2} C_{11}^{(2)}-
                 8 \bar{C}_{35}^{(2)}+6 \epsilon  C_{35}^{(2)}+
                 8 C_{23}^{(2)} k_1 \cdot k_2+
                 8 C_{33}^{(2)} k_1 \cdot k_2,
\end{array}
$$
$$
\begin{array} {lll}
f_{3}^{(2)} =& 4 m_{\tilde{g}}^{2} C_{0}^{(2)}-4 m_{\tilde{g}}^{2} C_{12}^{(2)}+
                8 \bar{C}_{24}^{(2)}-6 \epsilon  C_{24}^{(2)}-
                 8 \bar{C}_{36}^{(2)}+6 \epsilon  C_{36}^{(2)}\\
             &  + 8 k_1 \cdot k_2 C_{22}^{(2)}+
                 8 k_1 \cdot k_2 C_{34}^{(2)},
\end{array}
\eqno {(A.b.4)}
$$
$$
\begin{array} {lll}
f_{4}^{(2)} = &-4 m_{\tilde{g}}^{2} C_{0}^{(2)}-4 m_{\tilde{g}}^{2} C_{12}^{(2)}-
                8 \bar{C}_{24}^{(2)}+6 \epsilon  C_{24}^{(2)}-
                 8 \bar{C}_{36}^{(2)}+6 \epsilon  C_{36}^{(2)}+\\
             &+ 8 k_1 \cdot k_2 C_{12}^{(2)} +
                 8 k_1 \cdot k_2 C_{22}^{(2)} +
                 8 k_1 \cdot k_2 C_{23}^{(2)} +
                 8 k_1 \cdot k_2 C_{34}^{(2)},
\end{array}
$$
$$
\begin{array} {lll}
f_{5}^{(2)} = -8 C_{12}^{(2)}-24 C_{23}^{(2)}-16 C_{33}^{(2)}
\end{array}
$$
$$
\begin{array} {lll}
f_{6}^{(2)} = -8 C_{12}^{(2)}-16 C_{22}^{(2)}-8 C_{23}^{(2)}-16 C_{34}^{(2)}
\end{array}
$$
\par
   Similarly, the $gtt$ vertex functions are composed of left-handed
and right-handed contributions plus a counterterm:
(We define $a$ as color index of external gluon and $i,j$ as
colors of external top-quarks)
$$
\begin{array} {lll}
\Lambda_{\mu,(ij)}^{a}(p_1,p_2)= &
-\frac{g_{s} \hat{g}_{s}^{2}}{16 \pi^2} T^{a}_{ij}  \{ (2 C_{F}-C_{A})
( \Lambda_{\mu}^{(1L)}(p_1,p_2) P_{L}+
\Lambda_{\mu}^{(1R)}(p_1,p_2) P_{R})\\
& +C_{A}
(\Lambda_{\mu}^{(2L)}(p_1,p_2) P_{L}+\Lambda_{\mu}^{(2R)}(p_1,p_2) P_{R})\}\\
&+(m_{\tilde{t}_1} \rightarrow m_{\tilde{t}_2}, x_i \rightarrow y_i)+
 \Lambda_{\mu} ^{(CT)}\\
\end{array}
\eqno {(A.b.5)}
$$
The expressions of $\Lambda_{\mu}^{(n)},n=1L,1R,2L,2R$
are given as following:
$$
\begin{array} {lll}
\Lambda_{\mu}^{(n)}(p_1,p_2)= &h_{1}^{(n)} \gamma_{\mu}+
                   h_{2}^{(n)}  p_{1 \mu}+
                   h_{3}^{(n)}  p_{2 \mu}+
                   h_{4}^{(n)}  \rlap/p_{1} p_{1 \mu}\\
                   &+h_{5}^{(n)}  \rlap/p_{1} p_{2 \mu}+
                   h_{6}^{(n)}  \rlap/p_{2} p_{1 \mu}+
                   h_{7}^{(n)}  \rlap/p_{2} p_{2 \mu}+
                   h_{8}^{(n)} \gamma_{\mu} \rlap/p_{1}\\
                 &+h_{9}^{(n)} \gamma_{\mu} \rlap/p_{2}+
                  +h_{10}^{(n)} \gamma_{\mu} \rlap/p_{1} \rlap/p_{2}

\end{array}
\eqno {(A.b.6)}
$$
\par
 We define
$$
\begin{array} {lll}
C_{0}^{(3)},C_{ij}^{(3)}=C_{0},
 C_{ij}[-p_1,-p_2,m_{\tilde{t}_{1}},m_{\tilde{g}},m_{\tilde{t}_{1}}]
\end{array}
$$
$$
\begin{array} {lll}
  C_{0}^{(4)},C_{ij}^{(4)}= C_{0},
 C_{ij}[-p_1,-p_2,m_{\tilde{g}},m_{\tilde{t}_{1}},m_{\tilde{g}}]
\end{array}
$$
Then we can get $h_{i}^{(n)}$ as follows: (i=1,2,..., 10)
$$
\begin{array} {lll}
h_{1}^{(1L)}=-2 x_{2} x_{4} C_{24}^{(3)}
\end{array}
$$
$$
\begin{array} {lll}
h_{2}^{(1L)}=x_{2} x_{3} m_{\tilde{g}} (C_{0}^{(3)} +2 C_{11}^{(3)} )
\end{array}
$$
$$
\begin{array} {lll}
h_{3}^{(1L)}=x_{2} x_{3} m_{\tilde{g}} (C_{0}^{(3)} +2 C_{12}^{(3)} )
\end{array}
$$
$$
\begin{array} {lll}
h_{4}^{(1L)}=x_{2} x_{4} (C_{0}^{(3)} +3 C_{11}^{(3)} +2 C_{21}^{(3)} )
\end{array}
\eqno {(A.b.7)}
$$
$$
\begin{array} {lll}
h_{5}^{(1L)}=x_{2} x_{4} (C_{0}^{(3)} +C_{11}^{(3)} +2 C_{12}^{(3)} +2 C_{23}^{(3)} )
\end{array}
$$
$$
\begin{array} {lll}
h_{6}^{(1L)}=x_{2} x_{4} (C_{12}^{(3)} +2 C_{23}^{(3)} )
\end{array}
$$
$$
\begin{array} {lll}
h_{7}^{(1L)}=x_{2} x_{4} (C_{12}^{(3)} +2 C_{22}^{(3)} )
\end{array}
$$
$$
\begin{array} {lll}
h_{8}^{(1L)}=h_{9}^{(1L)}=h_{10}^{(1L)}=0
\end{array}
$$
and
$$
\begin{array} {lll}
h_{1}^{(2L)}=&x_{2} x_{4} (-m_{\tilde{g}}^{2} C_{0}^{(4)}-2 C_{24}^{(4)}+
              \epsilon  C_{24}^{(4)})\\
             &+x_{2} x_{4} p_{1}^{2} (C_{11}^{(4)}+C_{21}^{(4)})
              +2 x_{2} x_{4} p_{1} \dot p_{2} (C_{12}^{(4)}+C_{23}^{(4)}) \\
             &+ x_{2} x_{4} p_{2}^{2} (C_{12}^{(4)}+C_{22}^{(4)})
\end{array}
$$
$$
\begin{array} {lll}
h_{2}^{(2L)}=2 x_{2} x_{3} m_{\tilde{g}}^{2} C_{11}^{(4)}
\end{array}
$$
$$
\begin{array} {lll}
h_{3}^{(2L)}=2 x_{2} x_{3} m_{\tilde{g}}^{2} C_{12}^{(4)}
\end{array}
$$
$$
\begin{array} {lll}
h_{4}^{(2L)}=-2 x_{2} x_{4}(C_{11}^{(4)}+C_{21}^{(4)})
\end{array}
\eqno {(A.b.8)}
$$
$$
\begin{array} {lll}
h_{5}^{(2L)}=-2 x_{2} x_{4} (C_{12}^{(4)}+C_{23}^{(4)})
\end{array}
$$
$$
\begin{array} {lll}
h_{6}^{(2L)}=-2 x_{2} x_{4} (C_{11}^{(4)}+C_{23}^{(4)})
\end{array}
$$
$$
\begin{array} {lll}
h_{7}^{(2L)}=-2 x_{2} x_{4} (C_{12}^{(4)}+C_{22}^{(4)})
\end{array}
$$
$$
\begin{array} {lll}
h_{8}^{(2L)}=h_{9}^{(2L)}=x_{2} x_{3} m_{\tilde{g}}^{2} C_{0}^{(4)}
\end{array}
$$
$$
\begin{array} {lll}
h_{10}^{(2L)}=x_{2} x_{4} (C_{11}^{(4)}-C_{12}^{(4)})
\end{array}
$$
$h_{i}^{(1R)}$ and $h_{i}^{(2R)}$
can be obtained by exchanging $x_{1} \leftrightarrow x_{2}$
and $x_{3} \leftrightarrow x_{4}$ in $h_{i}^{(1L)}$ and $h_{i}^{(2L)}$.
($i=1,2,...,10$)
\par
The counter terms are given by:
$$
\begin{array} {lll}
\Lambda_{\mu}^{(CT)}= -C_{F} \frac{g_{s}}{2} T^{a}_{ij}
\gamma _{\mu} \left[ (\delta Z_{L}+ \delta Z_{L}^{\dagger}) P_{L}+
(\delta Z_{R}+ \delta Z_{R}^{\dagger}) P_{R}\right]
\end{array}
\eqno {(A.b.10)}
$$
The wave function renormalization constants can be obtained from
Eq.(3.c.12) and Eq.(3.c.13).
\par
C. Box corrections:
\par
Finally, we list the four form factors $F_{\mu\nu}^{ti}$ as given
in Eq.(3.e.1) in terms of Passarino-Veltmann functions. First,
we define $F_{k}^{tiL}$ and $F_{k}^{tiR}$ by:
$$
\begin{array} {lll}
   F_{\mu\nu}^{(ti)} ~ =&~ \frac{i \hat{g}_{s}^{2}}{16 \pi^{2}} P_{R}[
   \gamma _{\mu } \gamma _{\nu }F_{1}^{(tiR)}+
   \gamma _{\nu } \gamma _{\mu }F_{2}^{(tiR)}\\
 &+p_{1 \nu }\gamma _{\mu }F_{3}^{(tiR)} +
   p_{2 \nu } \gamma _{\mu }F_{4}^{(tiR)}\\
 &+p_{1 \mu }\gamma _{\nu } F_{5}^{(tiR)} +
   p_{2 \mu }\gamma _{\nu } F_{6}^{(tiR)}\\
 &+\gamma_{\mu} \gamma _{\nu } \rlap/k_{1} F_{7}^{(tiR)} +
   \gamma_{\nu} \gamma _{\mu } \rlap/k_{1} F_{8}^{(tiR)}\\
 &+\rlap/k_{1} p_{1 \mu} p_{2 \nu } F_{9}^{(tiR)} +
   \rlap/k_{1} p_{2 \mu} p_{1 \nu } F_{10}^{(tiR)} \\
 &+\gamma_{\mu} \rlap/k_{1} p_{1 \nu } F_{11}^{(tiR)} +
   \gamma_{\mu} \rlap/k_{1} p_{2 \nu } F_{12}^{(tiR)} \\
 &+\gamma_{\nu} \rlap/k_{1} p_{1 \mu } F_{13}^{(tiR)} +
   \gamma_{\nu} \rlap/k_{1} p_{2 \mu } F_{14}^{(tiR)} \\
 &+p_{1 \mu} p_{2 \nu } F_{15}^{(tiR)} +
   p_{1 \mu} p_{1 \nu } F_{16}^{(tiR)} \\
 &+p_{2 \mu} p_{1 \nu } F_{17}^{(tiR)} +
   \rlap/k_{1} p_{1 \mu} p_{1 \nu } F_{18}^{(tiR)}\\
 &+\rlap/k_{1} p_{2 \mu} p_{2 \nu } F_{19}^{(tiR)} +
   p_{2 \mu} p_{2 \nu } F_{20}^{(tiR)} ]\\
 &+(P_{R} \rightarrow P_{L}, F_{k}^{(tiR)} \rightarrow F_{k}^{(tiL)} ),
  (k=1 \sim 20,~i=1 \sim 4).
\end{array}
\eqno{(A.c.1)}
$$
In the following we only give the expressions of $F_{k}^{tiR}
(k=1,2,...,20~and ~i=1 \sim 4)$. The expressions
of $F_{k}^{tiL}$ can be obtained from $F_{k}^{tiR}$ by exchanging
$x_{1} \leftrightarrow x_{2}$ and $x_{3} \leftrightarrow x_{4}$.
Furthermore, the form factors in the u-channel are given by
$$
\begin{array} {lll}
F_{\mu\nu}^{(ui)}(k_1,k_2,p_1,p_2)=F_{\nu\mu}^{(ti)}(k_2,k_1,p_1,p_2)
\end{array}
\eqno{(A.c.2)}
$$
The expressions of $F_{k}^{(t1 R)}~(k=1 \sim 20)$ are given as below:
$$
\begin{array} {lll}
F_{1}^{(t1 R)}~ &=~F_{2}^{(t1 R)} \\
            &=-2 x_{1} x_{4} m_{\tilde{g}} D_{27}^{(1)}
             +2 x_{2} x_{4} m_{t} D_{311}^{(1)} \\
            &~+2 (x_{1} x_{3}-x_{2} x_{4}) m_{t} D_{313}^{(1)} ,
\end{array}
\eqno{(A.c.3)}
$$
$$
\begin{array} {lll}
F_{3}^{(t1 R)}~ =~4 x_{1} x_{3} (D_{311}^{(1)}-D_{312}^{(1)}) ,
\end{array}
$$
$$
\begin{array} {lll}
F_{4}^{(t1 R)}~ =~ - 4 x_{1} x_{3} (D_{27}^{(1)}+D_{312}^{(1)}) ,
\end{array}
$$
$$
\begin{array} {lll}
F_{5}^{(t1 R)}~ =~ 4 x_{1} x_{3}(D_{27}^{(1)}+D_{311}^{(1)}-D_{313}^{(1)}) ,
\end{array}
$$
$$
\begin{array} {lll}
F_{6}^{(t1 R)}~ =~ - 4 x_{1} x_{3} D_{313}^{(1)} ,
\end{array}
$$
$$
\begin{array} {lll}
F_{7}^{(t1 R)}~ =F_{8}^{(t1 R)}~= 2 x_{1}
x_{3}(D_{313}^{(1)}-D_{312}^{(1)}) ,
\end{array}
$$
$$
\begin{array} {lll}
F_{9}^{(t1 R)}~ =&~ 4 x_{1} x_{3} (D_{13}^{(1)}-D_{12}^{(1)}
                       -D_{22}^{(1)}-D_{23}^{(1)}-D_{24}^{(1)} \\
                       &+D_{25}^{(1)}+2 D_{26}^{(1)}-D_{36}^{(1)}
                       +D_{38}^{(1)}-D_{39}^{(1)}+D_{310}^{(1)}),
\end{array}
$$
$$
\begin{array} {lll}
F_{10}^{(t1 R)}~ =~ 4 x_{1} x_{3} (D_{37}^{(1)}
                         -D_{39}^{(1)}+D_{38}^{(1)}-D_{310}^{(1)}) ,
\end{array}
$$
$$
\begin{array} {lll}
F_{11}^{(t1 R)} =F_{12}^{(t1 R)}=
                F_{13}^{(t1 R)}= F_{14}^{(t1 R)}= 0 ,
\end{array}
$$
$$
\begin{array} {lll}
F_{15}^{(t1 R)}~ =&~ 4 x_{1} x_{3} m_{t} (D_{13}^{(1)}-D_{23}^{(1)}
                  +D_{25}^{(1)}+D_{26}^{(1)}-D_{39}^{(1)}+D_{310}^{(1)}) \\
                      &-4 x_{1} x_{4} m_{\tilde{g}} (D_{0}^{(1)}+D_{11}^{(1)}
                      +D_{12}^{(1)}-D_{13}^{(1)}+D_{24}^{(1)}-D_{26}^{(1)}) \\
                      &+4 x_{2} x_{4} m_{t} (D_{11}^{(1)}-D_{13}^{(1)}
                      +D_{21}^{(1)}+D_{23}^{(1)}+D_{24}^{(1)}-2 D_{25}^{(1)}\\
                      &-D_{26}^{(1)}+D_{34}^{(1)}+D_{39}^{(1)}-2 D_{310}^{(1)}) ,
\end{array}
$$
$$
\begin{array} {lll}
F_{16}^{(t1 R)}~ =&~ 4 x_{1} x_{3} m_{t} (-D_{25}^{(1)}
+D_{26}^{(1)}-D_{35}^{(1)}+D_{37}^{(1)}-D_{39}^{(1)}+D_{310}^{(1)})\\
         &+4 x_{1} x_{4} m_{\tilde{g}} (D_{11}^{(1)}-D_{12}^{(1)}
         +D_{21}^{(1)}-D_{24}^{(1)}-D_{25}^{(1)}+D_{26}^{(1)})\\
         &-4 x_{2} x_{4} m_{t} (D_{21}^{(1)}-D_{24}^{(1)}-D_{25}^{(1)}
         +D_{26}^{(1)}+D_{31}^{(1)} \\
         &-D_{34}^{(1)}-2 D_{35}^{(1)}+D_{37}^{(1)}-D_{39}^{(1)}+2 D_{310}^{(1)}) ,
\end{array}
$$
$$
\begin{array} {lll}
F_{17}^{(t1 R)}~ =&~ 4 x_{1} x_{3} m_{t} (D_{37}^{(1)}-D_{39}^{(1)})
                   - 4 x_{1} x_{4} m_{\tilde{g}} (D_{25}^{(1)}-D_{26}^{(1)}) \\
                  &+ 4 x_{2} x_{4} m_{t} (D_{35}^{(1)}-D_{37}^{(1)}+D_{39}^{(1)}-D_{310}^{(1)}) ,
\end{array}
$$
$$
\begin{array} {lll}
F_{18}^{(t1 R)}~ =&~ 4 x_{1} x_{3}
(-D_{22}^{(1)}+D_{24}^{(1)}-D_{25}^{(1)}+D_{26}^{(1)}+D_{34}^{(1)} \\
                        &-D_{35}^{(1)}-D_{36}^{(1)}+D_{37}^{(1)}+D_{38}^{(1)}-D_{39}^{(1)}) ,
\end{array}
$$
$$
\begin{array} {lll}
F_{19}^{(t1 R)}~ =~ 4 x_{1} x_{3}
(-D_{23}^{(1)}+D_{26}^{(1)}-D_{39}^{(1)}+D_{38}^{(1)}) ,
\end{array}
$$
$$
\begin{array} {lll}
F_{20}^{(t1 R)}~ =&~ 4 x_{1} x_{3} m_{t}
(-D_{23}^{(1)}-D_{39}^{(1)})+ 4 x_{1} x_{4} m_{\tilde{g}}
                         (D_{13}^{(1)}+D_{26}^{(1)}) \\
                        & +4 x_{2} x_{4} m_{t} (D_{23}^{(1)}-D_{25}^{(1)}
                        +D_{39}^{(1)}-D_{310}^{(1)}) ,
\end{array}
$$
where we denote $D_{i}^{(1)},D_{ij}^{(1)},D_{ijk}^{(1)}=
D_{i},D_{ij},D_{ijk} [-p_{1},k_{1},k_{2},m_{\tilde{g}},m_{\tilde{t_{1}}},m_{\tilde{t_{1}}},
m_{\tilde{t_{1}}}]$.
\par

The expressions of $F_{k}^{(t2 R)}~(k=1 \sim 20)$ are as follows:
$$
\begin{array} {lll}
F_{1}^{(t2 R)}~ =~ 2 x_{1} x_{3} m_{t} (D_{27}^{(2)}+D_{313}^{(2)})+
     2 x_{1} x_{4} m_{\tilde{g}} D_{27}^{(2)}- 2 x_{2} x_{4} m_{t} (D_{27}^{(2)}+D_{311}^{(2)}),
\end{array}
\eqno{(A.c.4)}
$$
$$
\begin{array} {lll}
F_{2}^{(t2 R)}~ =~ 2 x_{1} x_{3} m_{t} D_{313}^{(2)}+
     2 x_{1} x_{4} m_{\tilde{g}} D_{27}^{(2)}- 2 x_{2} x_{4} m_{t} D_{311}^{(2)},
\end{array}
$$
$$
\begin{array} {lll}
F_{3}^{(t2 R)}~ =~ 4 x_{1} x_{3} (-D_{27}^{(2)}-D_{311}^{(2)}+D_{312}^{(2)}),
\end{array}
$$
$$
\begin{array} {lll}
F_{4}^{(t2 R)}~ =~ 4 x_{1} x_{3} (D_{312}^{(2)}-D_{313}^{(2)}),
\end{array}
$$
$$
\begin{array} {lll}
F_{5}^{(t2 R)}~ =&~ 8 x_{1} x_{3} (D_{27}^{(2)}+D_{311}^{(2)}) +
2 x_{1} x_{3} m_{\tilde{g}}^{2} (D_{0}^{(2)}+D_{11}^{(2)})\\
&+2 x_{1} x_{3} m_{t}^{2} (-D_{11}^{(2)}-D_{13}^{(2)}-2 D_{21}^{(2)}-D_{23}^{(2)}
-D_{25}^{(2)}-D_{31}^{(2)}-D_{37}^{(2)})\\
&+2 (x_{2} x_{3}+x_{1} x_{4}) m_{\tilde{g}} m_{t} (D_{0}^{(2)}+D_{11}^{(2)})-
  2 x_{2} x_{4} m_{t}^{2} (D_{11}^{(2)}-D_{13}^{(2)}+D_{21}^{(2)}-D_{25}^{(2)})\\
&+4 x_{1} x_{3} k_{1} \cdot p_{1} (D_{12}^{(2)}+2 D_{24}^{(2)}+D_{34}^{(2)})\\
&+4 x_{1} x_{3} k_{1} \cdot p_{2} (D_{13}^{(2)}+D_{25}^{(2)}+D_{26}^{(2)}+D_{310}^{(2)})-
4 x_{1} x_{3} p_{1} \cdot p_{2} (D_{13}^{(2)}+2 D_{25}^{(2)}+ D_{35}^{(2)}) ,
\end{array}
$$
$$
\begin{array} {lll}
F_{6}^{(t2 R)}~ =&~
2 x_{1} x_{3} m_{\tilde{g}}^{2} D_{13}^{(2)}\\
&-2 x_{1} x_{3} m_{t}^{2} (D_{23}^{(2)}+D_{25}^{(2)}+D_{33}^{(2)}+D_{35}^{(2)})+
2 (x_{2} x_{3}+x_{1} x_{4}) m_{\tilde{g}} m_{t} D_{13}^{(2)}\\
&+2 x_{2} x_{4} m_{t}^{2} (D_{23}^{(2)}-D_{25}^{(2)})+
4 x_{1} x_{3} k_{1} \cdot p_{1} (D_{26}^{(2)}+D_{310}^{(2)})\\
&+4 x_{1} x_{3} k_{1} \cdot p_{2} (D_{23}^{(2)}+D_{39}^{(2)})-
4 x_{1} x_{3} p_{1} \cdot p_{2} (D_{23}^{(2)}+D_{37}^{(2)})\\
&+ 8 x_{1} x_{3} D_{313}^{(2)},
\end{array}
$$
$$
\begin{array} {lll}
F_{7}^{(t2 R)}~ =~ 2 x_{1} x_{3} (D_{27}^{(2)}+D_{312}^{(2)}),
\end{array}
$$
$$
\begin{array} {lll}
F_{8}^{(t2 R)}~ =~ 2 x_{1} x_{3} D_{312}^{(2)},
\end{array}
$$
$$
\begin{array} {lll}
F_{9}^{(t2 R)}~ =~ 4 x_{1} x_{3} (D_{12}^{(2)}-D_{13}^{(2)}+D_{22}^{(2)}+
      D_{24}^{(2)}-D_{25}^{(2)}-D_{26}^{(2)}+D_{36}^{(2)}-D_{310}^{(2)}) ,
\end{array}
$$
$$
\begin{array} {lll}
F_{10}^{(t2 R)}~ =~ 4 x_{1} x_{3} (D_{38}^{(2)}-D_{310}^{(2)}) ,
\end{array}
$$
$$
\begin{array} {lll}
F_{11}^{(t2 R)} =F_{12}^{(t2 R)}= 0 ,
\end{array}
$$
$$
\begin{array} {lll}
F_{13}^{(t2 R)}~ =&~ 2 x_{1} x_{3} m_{t} (-D_{12}^{(2)}+D_{13}^{(2)}-D_{24}^{(2)}+D_{25}^{(2)})+
     2 x_{1} x_{4} m_{\tilde{g}} (D_{0}^{(2)}+D_{11}^{(2)})\\
     &-2 x_{2} x_{4} m_{t} (D_{11}^{(2)}-D_{12}^{(2)}+D_{21}^{(2)}-D_{24}^{(2)}) ,
\end{array}
$$
$$
\begin{array} {lll}
F_{14}^{(t2 R)}~ =~ 2 x_{1} x_{3} m_{t} (D_{23}^{(2)}-D_{26}^{(2)})+
     2 x_{1} x_{4} m_{\tilde{g}} D_{13}^{(2)}-
     2 x_{2} x_{4} m_{t} (D_{25}^{(2)}-D_{26}^{(2)}) ,
\end{array}
$$
$$
\begin{array} {lll}
F_{15}^{(t2 R)}~ =&~ 4 x_{1} x_{3} m_{t} (D_{12}^{(2)}-D_{13}^{(2)}-D_{23}^{(2)}+
D_{24}^{(2)}-D_{25}^{(2)}+D_{26}^{(2)}-D_{37}^{(2)}+D_{310}^{(2)})\\
     &+4 x_{1} x_{4} m_{\tilde{g}} (D_{12}^{(2)}-D_{13}^{(2)}+D_{24}^{(2)}-D_{25}^{(2)})\\
     &-4 x_{2} x_{4} m_{t} (D_{12}^{(2)}-D_{13}^{(2)}+2 D_{24}^{(2)}-2 D_{25}^{(2)}+D_{34}^{(2)}-D_{35}^{(2)}) ,
\end{array}
$$
$$
\begin{array} {lll}
F_{16}^{(t2 R)}~ =&~ 4 x_{1} x_{3} m_{t} (D_{12}^{(2)}-D_{13}^{(2)}+D_{24}^{(2)}-2 D_{25}^{(2)}+
D_{26}^{(2)}-D_{35}^{(2)}+D_{310}^{(2)})\\
   &+4 x_{1} x_{4} m_{\tilde{g}} (-D_{0}^{(2)}-2 D_{11}^{(2)}+D_{12}^{(2)}-D_{21}^{(2)}+D_{24}^{(2)})\\
   &+4 x_{2} x_{4} m_{t} (D_{11}^{(2)}-D_{12}^{(2)}+2 D_{21}^{(2)}-2 D_{24}^{(2)}+D_{31}^{(2)}-D_{34}^{(2)}) ,
\end{array}
$$
$$
\begin{array} {lll}
F_{17}^{(t2 R)}~ =&~ 4 x_{1} x_{3} m_{t} (-D_{23}^{(2)}+D_{26}^{(2)}-D_{37}^{(2)}+
  D_{39}^{(2)})+
   4 x_{1} x_{4} m_{\tilde{g}} (-D_{13}^{(2)}-D_{25}^{(2)}+D_{26}^{(2)})\\
   &+ 4 x_{2} x_{4} m_{t} (D_{25}^{(2)}-D_{26}^{(2)}+D_{35}^{(2)}-D_{310}^{(2)}) ,
\end{array}
$$
$$
\begin{array} {lll}
F_{18}^{(t2 R)}~ =~ 4 x_{1} x_{3} (D_{22}^{(2)}-D_{24}^{(2)}-D_{34}^{(2)}
                      +D_{36}^{(2)}),
\end{array}
$$
$$
\begin{array} {lll}
F_{19}^{(t2 R)}~ =~ 4 x_{1} x_{3} (-D_{23}^{(2)}+D_{26}^{(2)}
                       +D_{38}^{(2)}-D_{39}^{(2)}),
\end{array}
$$
$$
\begin{array} {lll}
F_{20}^{(t2 R)}~ =&~ 4 x_{1} x_{3} m_{t} (-D_{23}^{(2)}+D_{26}^{(2)}-D_{33}^{(2)}+
  D_{39}^{(2)})+ 4 x_{1} x_{4} m_{\tilde{g}} (-D_{23}^{(2)}+D_{26}^{(2)})\\
   &+4 x_{2} x_{4} m_{t} (D_{23}^{(2)}-D_{26}^{(2)}+D_{37}^{(2)}-D_{310}^{(2)}) ,
\end{array}
$$
where $D_{i}^{(2)},D_{ij}^{(2)},D_{ijk}^{(2)} ~=~ D_{i},D_{ij},D_{ijk}
[-p_{1},k_{1},-p_{2},m_{\tilde{g}},m_{\tilde{t_{1}}},m_{\tilde{t_{1}}},m_{\tilde{g}}]$.

\par
The expressions for $F_{k}^{(t3 R)}~(k=1 \sim 20)$ are written as:
$$
\begin{array} {lll}
F_{1}^{(t3 R)}~ =&
       2 x_{1} x_{3} m_{t} (D_{27}^{(3)}+2 D_{313}^{(3)})+
       x_{1} x_{3} m_{t} m_{\tilde{g}}^{2} (D_{0}^{(3)}+D_{13}^{(3)})\\
       &-x_{1} x_{3} m_{t}^{3} (D_{0}^{(3)}+2 D_{11}^{(3)}-D_{13}^{(3)}+D_{21}^{(3)}+2 D_{33}^{(3)}
       +2 D_{35}^{(3)}-2 D_{37}^{(3)})\\
       &+2 x_{1} x_{4} m_{\tilde{g}} D_{27}^{(3)}+
       x_{1} x_{4} m_{\tilde{g}}^{3} D_{0}^{(3)}\\
       &-x_{1} x_{4} m_{t}^{2} m_{\tilde{g}} (D_{0}^{(3)}+2 D_{11}^{(3)}-
       2 D_{13}^{(3)}+D_{21}^{(3)}+2 D_{23}^{(3)}-2 D_{25}^{(3)})\\
       &+2 x_{2} x_{4} m_{t} (D_{27}^{(3)}+2 D_{311}^{(3)}-2 D_{313}^{(3)}) +
       x_{2} x_{4} m_{t} m_{\tilde{g}}^{2} (D_{11}^{(3)}-D_{13}^{(3)}) \\
       &-x_{2} x_{4} m_{t}^{3} (D_{11}^{(3)}-D_{13}^{(3)}+2 D_{21}^{(3)}+2 D_{23}^{(3)}-4 D_{25}^{(3)}+
       D_{31}^{(3)}-2 D_{33}^{(3)}-3 D_{35}^{(3)}+4 D_{37}^{(3)})\\
       &+2 x_{1} x_{3} m_{t} k_{1} \cdot p_{1} (D_{12}^{(3)}-D_{13}^{(3)}-D_{23}^{(3)}+D_{24}^{(3)}
       +D_{33}^{(3)}-D_{37}^{(3)}-D_{39}^{(3)}+D_{310}^{(3)})\\
       &+2 x_{1} x_{4} m_{\tilde{g}} k_{1} \cdot p_{1} (D_{11}^{(3)}+D_{12}^{(3)}-
       2 D_{13}^{(3)}+D_{23}^{(3)}+D_{24}^{(3)}-D_{25}^{(3)}-D_{26}^{(3)})\\
       &+2 x_{2} x_{4} m_{t} k_{1} \cdot p_{1} (D_{11}^{(3)}-D_{13}^{(3)}+D_{21}^{(3)}+2 D_{23}^{(3)}\\
       &+D_{24}^{(3)}-3 D_{25}^{(3)}-D_{26}^{(3)}-D_{33}^{(3)}+D_{34}^{(3)}-D_{35}^{(3)}+2 D_{37}^{(3)}
        +D_{39}^{(3)}-2 D_{310}^{(3)})\\
       &+2 x_{1} x_{3} m_{t} k_{1} \cdot p_{2} (-D_{13}^{(3)}-D_{26}^{(3)}+D_{33}^{(3)}-D_{39}^{(3)})\\
       &+2 x_{1} x_{4} m_{\tilde{g}} k_{1} \cdot p_{2} (-D_{13}^{(3)}+D_{23}^{(3)}-D_{26}^{(3)})\\
       &+2 x_{2} x_{4} m_{t} k_{1} \cdot p_{2} (D_{23}^{(3)}-D_{25}^{(3)}-D_{33}^{(3)}+
       D_{37}^{(3)}+D_{39}^{(3)}-D_{310}^{(3)})\\
       &+2 x_{1} x_{3} m_{t} p_{1} \cdot p_{2} (D_{13}^{(3)}+D_{25}^{(3)}-D_{33}^{(3)}+D_{37}^{(3)})\\
       &+2 x_{1} x_{4} m_{\tilde{g}} p_{1} \cdot p_{2} (D_{13}^{(3)}-D_{23}^{(3)}+D_{25}^{(3)})\\
       &+2 x_{2} x_{4} m_{t} p_{1} \cdot p_{2}
       (-D_{23}^{(3)}+D_{25}^{(3)}+D_{33}^{(3)}+D_{35}^{(3)}-D_{37}^{(3)}),
\end{array}
\eqno{(A.c.5)}
$$
$$
\begin{array} {lll}
F_{2}^{(t3 R)}~ =~ - 2 x_{1} x_{4} m_{\tilde{g}} D_{27}^{(3)}
            -2 x_{1} x_{3} m_{t} D_{313}^{(3)}
            - 2 x_{2} x_{4} m_{t} (D_{27}^{(3)}+D_{311}^{(3)}-D_{313}^{(3)}),
\end{array}
$$
$$
\begin{array} {lll}
F_{3}^{(t3 R)}~ =&
     4 x_{1} x_{3} (D_{27}^{(3)}+2 D_{311}^{(3)}+D_{312}^{(3)}-3 D_{313}^{(3)})+
     2 x_{1} x_{3} m_{\tilde{g}}^{2} (-D_{0}^{(3)}+D_{11}^{(3)}-D_{13}^{(3)})\\
     &+2 x_{1} x_{3} m_{t}^{2} (D_{13}^{(3)}-D_{21}^{(3)}+2 D_{25}^{(3)}-D_{31}^{(3)}+2 D_{33}^{(3)}+
     3 D_{35}^{(3)}-4 D_{37}^{(3)})\\
     &-2 x_{2} x_{3} m_{t} m_{\tilde{g}} D_{0}^{(3)}-
     2 x_{1} x_{4} m_{t} m_{\tilde{g}} D_{0}^{(3)}-
     2 x_{2} x_{4} m_{t}^{2} (D_{0}^{(3)}+D_{11}^{(3)})\\
     &+4 x_{1} x_{3} k_{1} \cdot p_{1} (D_{23}^{(3)}+D_{24}^{(3)}-D_{25}^{(3)}
     -D_{26}^{(3)}\\
     &-D_{33}^{(3)}+D_{34}^{(3)}-D_{35}^{(3)}+2 D_{37}^{(3)}+D_{39}^{(3)}
       -2 D_{310}^{(3)})\\
     &+4 x_{1} x_{3} k_{1} \cdot p_{2} (-D_{25}^{(3)}+D_{26}^{(3)}-D_{33}^{(3)}+D_{37}^{(3)}+
      D_{39}^{(3)}-D_{310}^{(3)})\\
     &+4 x_{1} x_{3} p_{1} \cdot p_{2} (D_{33}^{(3)}+D_{35}^{(3)}-2 D_{37}^{(3)}) ,
\end{array}
$$
$$
\begin{array} {lll}
F_{4}^{(t3 R)}~ =&
     4 x_{1} x_{3}  D_{312}^{(3)}-
     2 x_{1} x_{3} m_{\tilde{g}}^{2} D_{0}^{(3)}\\
     &+2 x_{1} x_{3} m_{t}^{2} (D_{11}^{(3)}+D_{21}^{(3)}+2 D_{23}^{(3)}-2 D_{25}^{(3)})\\
     &-2 x_{2} x_{3} m_{t} m_{\tilde{g}} D_{0}^{(3)}-
     2 x_{1} x_{4} m_{t} m_{\tilde{g}} D_{0}^{(3)}\\
     &-2 x_{2} x_{4} m_{t}^{2} (D_{0}^{(3)}+D_{11}^{(3)})+
     4 x_{1} x_{3} k_{1} \cdot p_{1} (-D_{23}^{(3)}+D_{25}^{(3)})\\
     &+4 x_{1} x_{3} k_{1} \cdot p_{2} (-D_{23}^{(3)}+D_{26}^{(3)})+
     4 x_{1} x_{3} p_{1} \cdot p_{2} (D_{23}^{(3)}-D_{25}^{(3)}) ,
\end{array}
$$
$$
\begin{array} {lll}
F_{5}^{(t3 R)}~ =&  4 x_{1} x_{3} (-D_{311}^{(3)}+D_{313}^{(3)})+
                      2 x_{1} x_{3} m_{\tilde{g}}^{2} D_{0}^{(3)}\\
                      &-2 x_{1} x_{3} m_{t}^{2} (D_{13}^{(3)}+D_{25}^{(3)})+
                      2 x_{2} x_{3} m_{t} m_{\tilde{g}} (D_{0}^{(3)}+D_{11}^{(3)}-
                      D_{13}^{(3)})\\
                      &+2 x_{1} x_{4} m_{t} m_{\tilde{g}} (D_{0}^{(3)}+D_{11}^{(3)}-
                      D_{13}^{(3)})\\
                      &+2 x_{2} x_{4} m_{t}^{2} (D_{0}^{(3)}+2 D_{11}^{(3)}-D_{13}^{(3)}+
                      D_{21}^{(3)}-D_{25}^{(3)})\\
                      &+4 x_{1} x_{3} k_{1} \cdot p_{2} (D_{25}^{(3)}-D_{26}^{(3)}),
\end{array}
$$
$$
\begin{array} {lll}
F_{6}^{(t3 R)}~ =&
                     -8 x_{1} x_{3} D_{313}^{(3)}-
                      2 x_{1} x_{3} m_{\tilde{g}}^{2} D_{13}^{(3)}\\
                      &+2 x_{1} x_{3} m_{t}^{2} (D_{25}^{(3)}+2 D_{33}^{(3)}+D_{35}^{(3)}-2 D_{37}^{(3)})-
                      2 x_{2} x_{3} m_{t} m_{\tilde{g}} D_{13}^{(3)}\\
                      &-2 x_{1} x_{4} m_{t} m_{\tilde{g}} D_{13}^{(3)}-
                      2 x_{2} x_{4} m_{t}^{2} (D_{13}^{(3)}+D_{25}^{(3)})\\
                     &+4 x_{1} x_{3} k_{1} \cdot p_{1} (D_{23}^{(3)}
                     -D_{25}^{(3)}- D_{33}^{(3)}+D_{37}^{(3)}+D_{39}^{(3)}
                     -D_{310}^{(3)})\\
                     &+4 x_{1} x_{3} k_{1} \cdot p_{2} (-D_{33}^{(3)}+D_{39}^{(3)})+
                     4 x_{1} x_{3} p_{1} \cdot p_{2} (D_{33}^{(3)}-D_{37}^{(3)}),
\end{array}
$$
$$
\begin{array} {lll}
F_{7}^{(t3 R)}~ =&~  -4 x_{1} x_{3} (D_{312}^{(3)}-D_{313}^{(3)})+
                      x_{1} x_{3} m_{\tilde{g}}^{2} (D_{0}^{(3)}-D_{12}^{(3)}+D_{13}^{(3)})\\
                      &+x_{1} x_{3} m_{t}^{2} (-D_{11}^{(3)}+D_{12}^{(3)}-D_{13}^{(3)}-D_{21}^{(3)}\\
                   &+2 D_{24}^{(3)}-2 D_{26}^{(3)}-2 D_{33}^{(3)}+D_{34}^{(3)}
                      -D_{35}^{(3)}+2 D_{37}^{(3)}
                      +2 D_{39}^{(3)}-2 D_{310}^{(3)})\\
                   &+(x_{2} x_{3}+x_{1} x_{4}) m_{t} m_{\tilde{g}} D_{0}^{(3)}
                    + x_{2} x_{4} m_{t}^{2} (D_{0}^{(3)}+D_{11}^{(3)})\\
                   &+2 x_{1} x_{3} k_{1} \cdot p_{1} (-D_{22}^{(3)}
                    -D_{23}^{(3)}+2 D_{26}^{(3)} + D_{33}^{(3)}-D_{36}^{(3)}
                    -D_{37}^{(3)}+D_{38}^{(3)}-2 D_{39}^{(3)}+2 D_{310}^{(3)})\\
                   &+2 x_{1} x_{3} k_{1} \cdot p_{2} (D_{33}^{(3)}+D_{38}^{(3)}-2 D_{39}^{(3)})\\
                   &+2 x_{1} x_{3} p_{1} \cdot p_{2} (D_{25}^{(3)}-D_{26}^{(3)}-
                    D_{33}^{(3)}+D_{37}^{(3)}+D_{39}^{(3)}-D_{310}^{(3)}),
\end{array}
$$
$$
\begin{array} {lll}
F_{8}^{(t3 R)}~ =~  2 x_{1} x_{3} (D_{27}^{(3)}+D_{312}^{(3)}-D_{313}^{(3)}),
\end{array}
$$
$$
\begin{array} {lll}
F_{9}^{(t3 R)}~ =~  4 x_{1} x_{3} (D_{22}^{(3)}+D_{23}^{(3)}-D_{25}^{(3)}-D_{26}^{(3)}
             +D_{36}^{(3)}-D_{38}^{(3)}+D_{39}^{(3)}-D_{310}^{(3)}),
\end{array}
$$
$$
\begin{array} {lll}
F_{10}^{(t3 R)}~ =~  4 x_{1} x_{3} (D_{25}^{(3)}-D_{26}^{(3)}
                   -D_{37}^{(3)}-D_{38}^{(3)}+D_{39}^{(3)}+D_{310}^{(3)}),
\end{array}
$$
$$
\begin{array} {lll}
F_{11}^{(t3 R)}~ =&~  2 x_{1} x_{3} m_{t} (-D_{25}^{(3)}+D_{26}^{(3)})+
                       2 x_{1} x_{4} m_{\tilde{g}} (-D_{11}^{(3)}
                       +D_{12}^{(3)})\\
                       &+2 x_{2} x_{4} m_{t} (-D_{11}^{(3)}+D_{12}^{(3)}
                       -D_{21}^{(3)}+ D_{24}^{(3)}+D_{25}^{(3)}
                       -D_{26}^{(3)}),
\end{array}
$$
$$
\begin{array} {lll}
F_{12}^{(t3 R)}~ =&~  2 x_{1} x_{3} m_{t} (D_{13}^{(3)}+D_{26}^{(3)})+
                       2 x_{1} x_{4} m_{\tilde{g}} D_{12}^{(3)}\\
                       &+2 x_{2} x_{4} m_{t} (D_{12}^{(3)}-D_{13}^{(3)}
                       +D_{24}^{(3)}-D_{26}^{(3)}),
\end{array}
$$
$$
\begin{array} {lll}
F_{13}^{(t3 R)}~ =&~  2 x_{1} x_{3} m_{t} (D_{12}^{(3)}-D_{13}^{(3)}+D_{24}^{(3)}-D_{26}^{(3)})+
                       2 x_{1} x_{4} m_{\tilde{g}} (D_{11}^{(3)}-D_{13}^{(3)})\\
                       &+2 x_{2} x_{4} m_{t} (D_{11}^{(3)}-D_{12}^{(3)}+D_{21}^{(3)}-D_{24}^{(3)}-
D_{25}^{(3)}+D_{26}^{(3)}),
\end{array}
$$
$$
\begin{array} {lll}
F_{14}^{(t3 R)}~ =&~  -2 x_{1} x_{3} m_{t} (D_{13}^{(3)}+D_{26}^{(3)})-
                       2 x_{1} x_{4} m_{\tilde{g}} D_{13}^{(3)}\\
                       &+2 x_{2} x_{4} m_{t} (-D_{25}^{(3)}+D_{26}^{(3)}),
\end{array}
$$
$$
\begin{array} {lll}
F_{15}^{(t3 R)}~ =&  -4 x_{1} x_{3} m_{t}
(D_{12}^{(3)}-D_{23}^{(3)}+D_{24}^{(3)}+D_{25}^{(3)}
-D_{39}^{(3)}+D_{310}^{(3)})\\
                       &+4 x_{1} x_{4} m_{\tilde{g}} (-D_{11}^{(3)}-D_{12}^{(3)}+D_{13}^{(3)}-
                       D_{24}^{(3)}+D_{26}^{(3)})\\
                       &+4 x_{2} x_{4} m_{t} (-D_{11}^{(3)}+D_{13}^{(3)}-D_{21}^{(3)}-D_{23}^{(3)}\\
                       &-D_{24}^{(3)}+2 D_{25}^{(3)}+D_{26}^{(3)}
                        -D_{34}^{(3)}-D_{39}^{(3)}+2 D_{310}^{(3)}),
\end{array}
$$
$$
\begin{array} {lll}
F_{16}^{(t3 R)}~ =&  -4 x_{1} x_{3} m_{t} (D_{12}^{(3)}-D_{13}^{(3)}+D_{24}^{(3)}
                       -D_{25}^{(3)}-D_{35}^{(3)}+D_{37}^{(3)}-D_{39}^{(3)}+ D_{310}^{(3)})\\
                       &+4 x_{1} x_{4} m_{\tilde{g}} (-D_{12}^{(3)}+D_{13}^{(3)}+D_{21}^{(3)}-
                       D_{24}^{(3)}-D_{25}^{(3)}+D_{26}^{(3)})\\
                       &+4 x_{2} x_{4} m_{t} (D_{21}^{(3)}-D_{24}^{(3)}- D_{25}^{(3)}+D_{26}^{(3)}\\
                       &+D_{31}^{(3)}-D_{34}^{(3)}-2 D_{35}^{(3)}+D_{37}^{(3)}-
                       D_{39}^{(3)}+2 D_{310}^{(3)}),
\end{array}
$$
$$
\begin{array} {lll}
F_{17}^{(t3 R)}~ =&~  4 x_{1} x_{3} m_{t} (D_{13}^{(3)}
                       +D_{26}^{(3)}-D_{37}^{(3)}+D_{39}^{(3)})\\
                       &+4 x_{1} x_{4} m_{\tilde{g}} (D_{13}^{(3)}
                       -D_{25}^{(3)}+D_{26}^{(3)})+
                       4 x_{2} x_{4} m_{t} (-D_{35}^{(3)}
                       +D_{37}^{(3)}-D_{39}^{(3)}+D_{310}^{(3)}),
\end{array}
$$
$$
\begin{array} {lll}
F_{18}^{(t3 R)}~ =&~  4 x_{1} x_{3} (D_{22}^{(3)}-D_{24}^{(3)}
                   +D_{25}^{(3)}-D_{26}^{(3)}-
                    D_{34}^{(3)}+D_{35}^{(3)}+D_{36}^{(3)}\\
                   & -D_{37}^{(3)}-D_{38}^{(3)}+D_{39}^{(3)}),
\end{array}
$$
$$
\begin{array} {lll}
F_{19}^{(t3 R)}~ =~  4 x_{1} x_{3} (D_{23}^{(3)}-D_{26}^{(3)}
                     -D_{38}^{(3)}+D_{39}^{(3)}),
\end{array}
$$
$$
\begin{array} {lll}
F_{20}^{(t3 R)}~ =&~  4 x_{1} x_{3} m_{t} (D_{13}^{(3)}+D_{23}^{(3)}
                       +D_{26}^{(3)}+D_{39}^{(3)})\\
                       &+4 x_{1} x_{4} m_{\tilde{g}} (D_{13}^{(3)}
                       +D_{26}^{(3)})+
                       4 x_{2} x_{4} m_{t} (-D_{23}^{(3)}+D_{25}^{(3)}
                       -D_{39}^{(3)}+D_{310}^{(3)}),
\end{array}
$$
where $D_{i}^{(3)},D_{ij}^{(3)},D_{ijk}^{(3)}~=~
D_{i},D_{ij},D_{ijk} [-p_{1},k_{1},k_{2},m_{\tilde{t_{1}}},m_{\tilde{g}},m_{\tilde{g}},
m_{\tilde{g}}]$.
\par
The $F_{k}^{(t4 R)}~(k=1 \sim 20)$ are written explicitly as:
$$
\begin{array} {lll}
F_{1}^{(t4 R)} &= F_{2}^{(t4 R)} \\
& =\frac{1}{2} ((x_{1} x_{3} m_{t} (C_{11}-C_{12})\\
                      &+x_{2} x_{4} m_{t} C_{12}-
                      x_{2} x_{3} m_{\tilde{g}} C_{0})[-p_1,p_1+p_2,
                      m_{\tilde{g}},m_{\tilde{t}_{1}},m_{\tilde{t}_{1}}])
\end{array}
\eqno{(A.c.6)}
$$
$$
\begin{array} {lll}
F_{i}^{(t4 R)}~ =~ 0, (i=3,4,...20).
\end{array}
$$

%\newpage

%\newpage

\begin{center}
{\large \bf Figure Captions}
\end{center}

\parindent=0pt

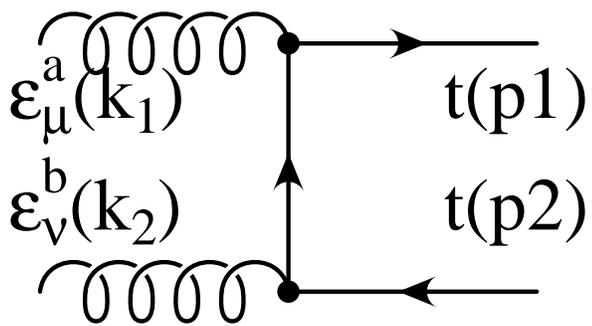
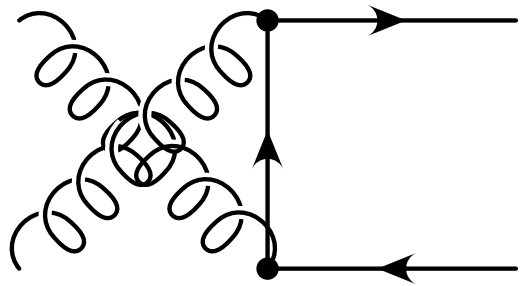
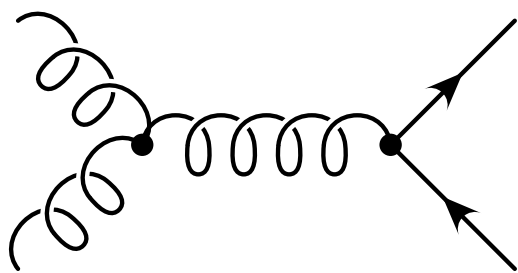
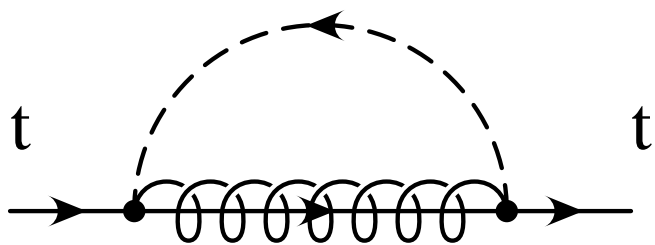
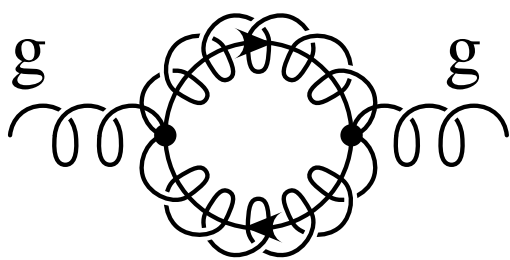
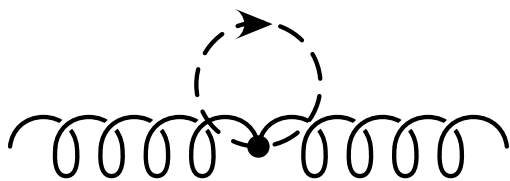
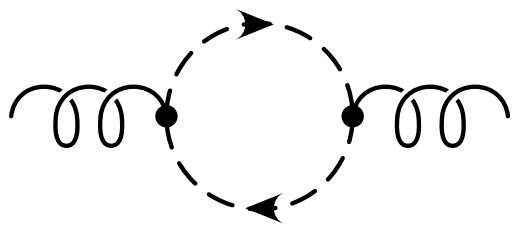
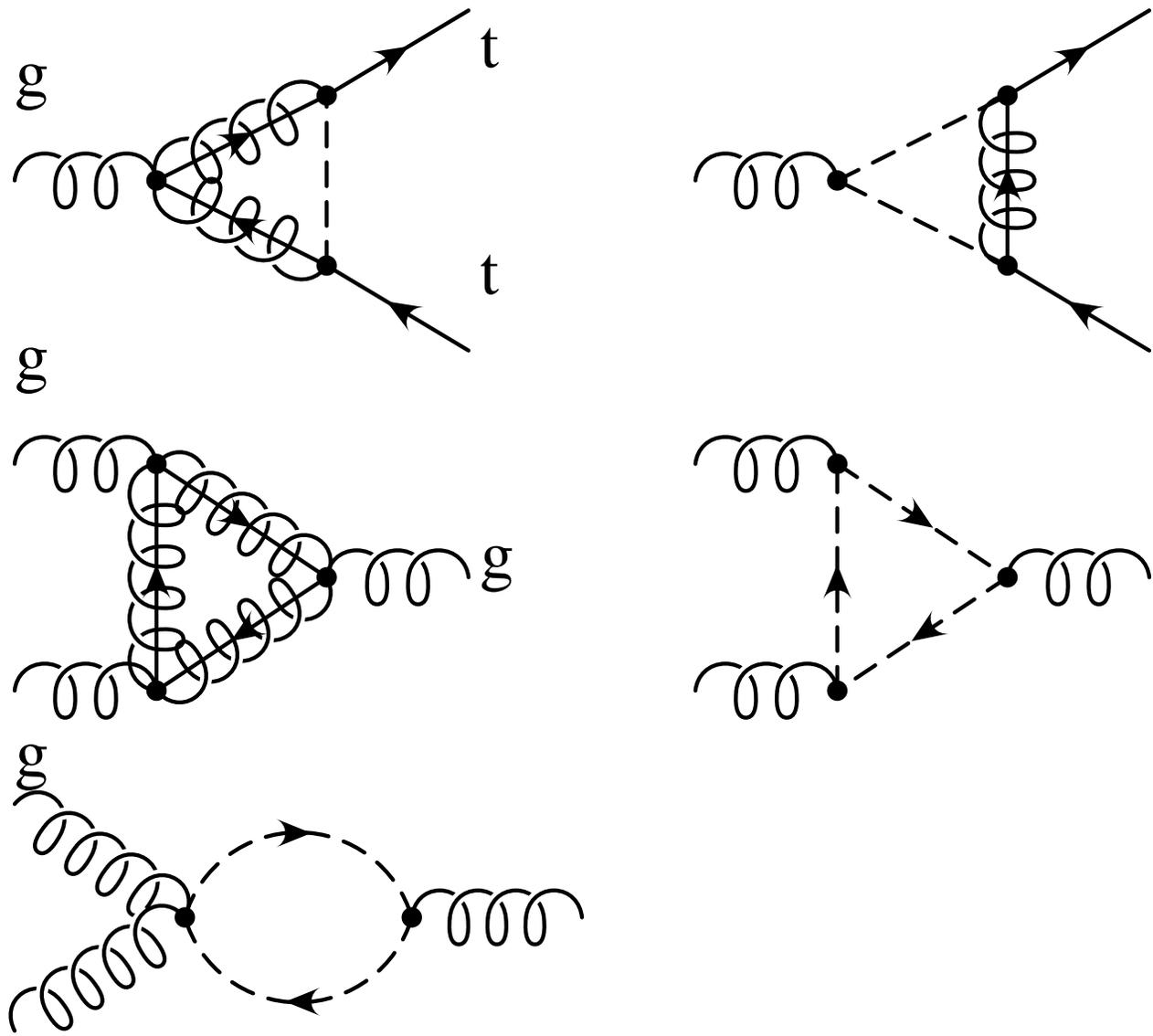
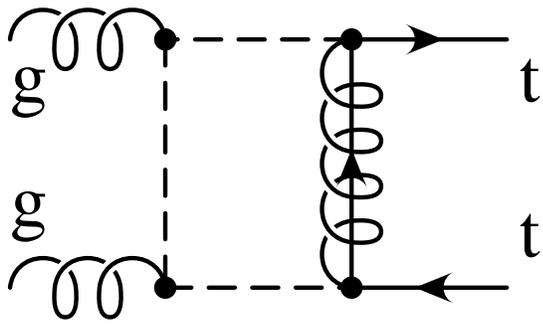
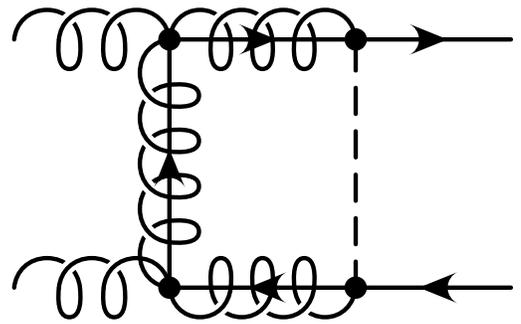
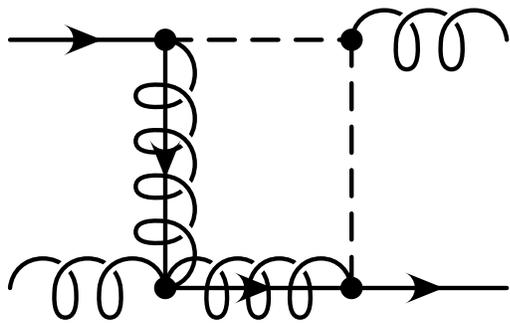
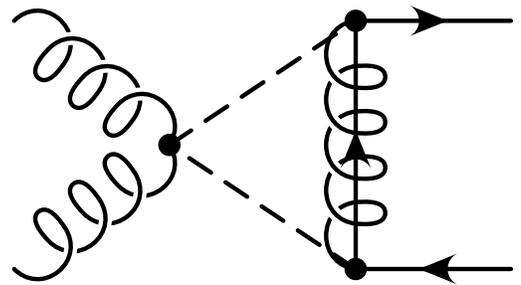
{\bf Fig.1} Feynman diagrams at
        the tree-level and one-loop level in the SUSY
        QCD for the $gg\rightarrow t \bar{t}$ subprocess.
Fig.1 (a): Tree level diagrams.
Fig.1 (b.1): Self-energy diagrams (for top-quark and gluon).
Fig.1 (b.2): Vertex diagrams (including tri-gluon and gluon-top-top
             interactions).
Fig.1 (b.3): Box diagrams (only t-channel).
Dashed lines represent $\tilde{t}_1,\tilde{t}_2$ in Fig.1 (b).

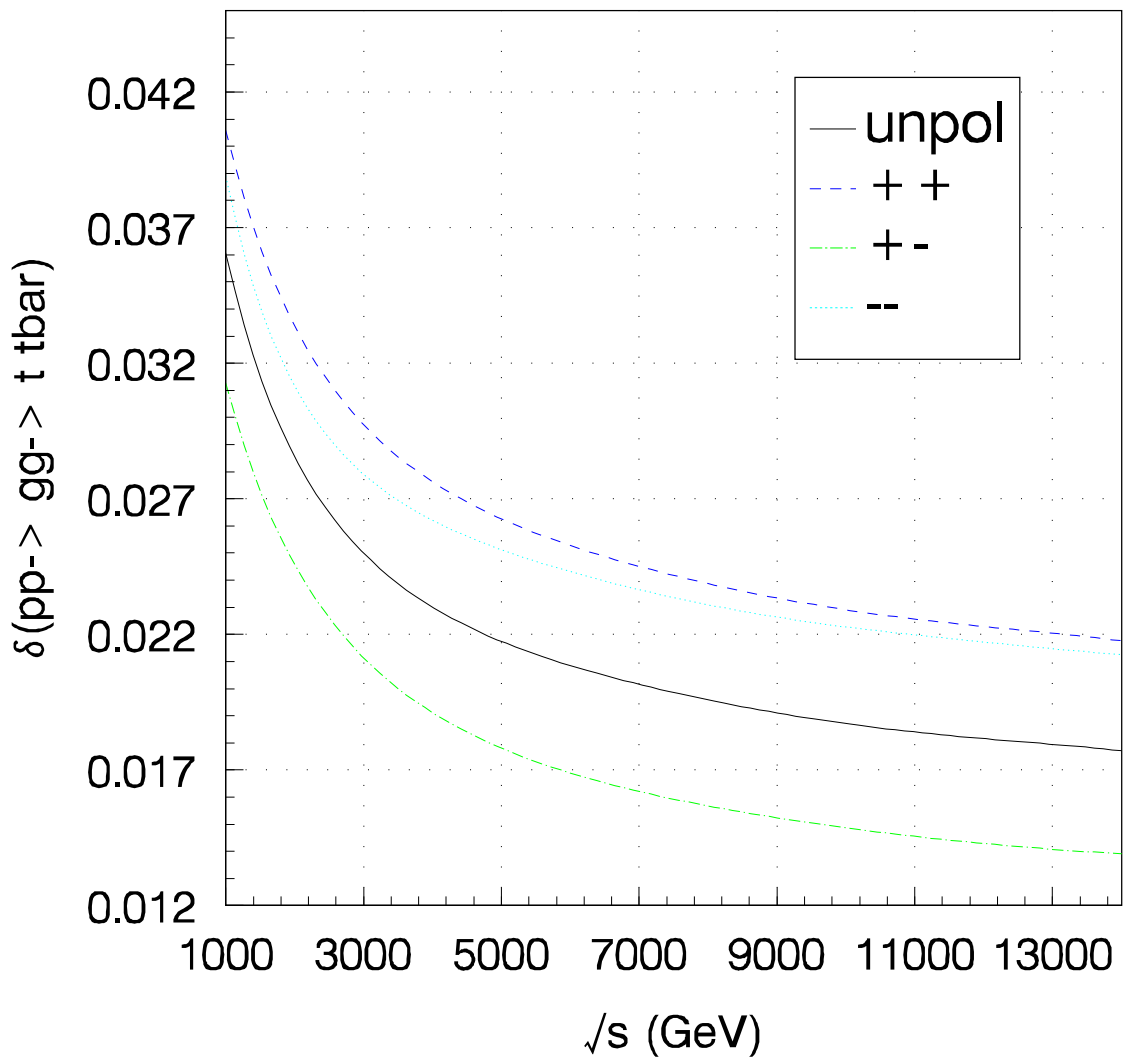
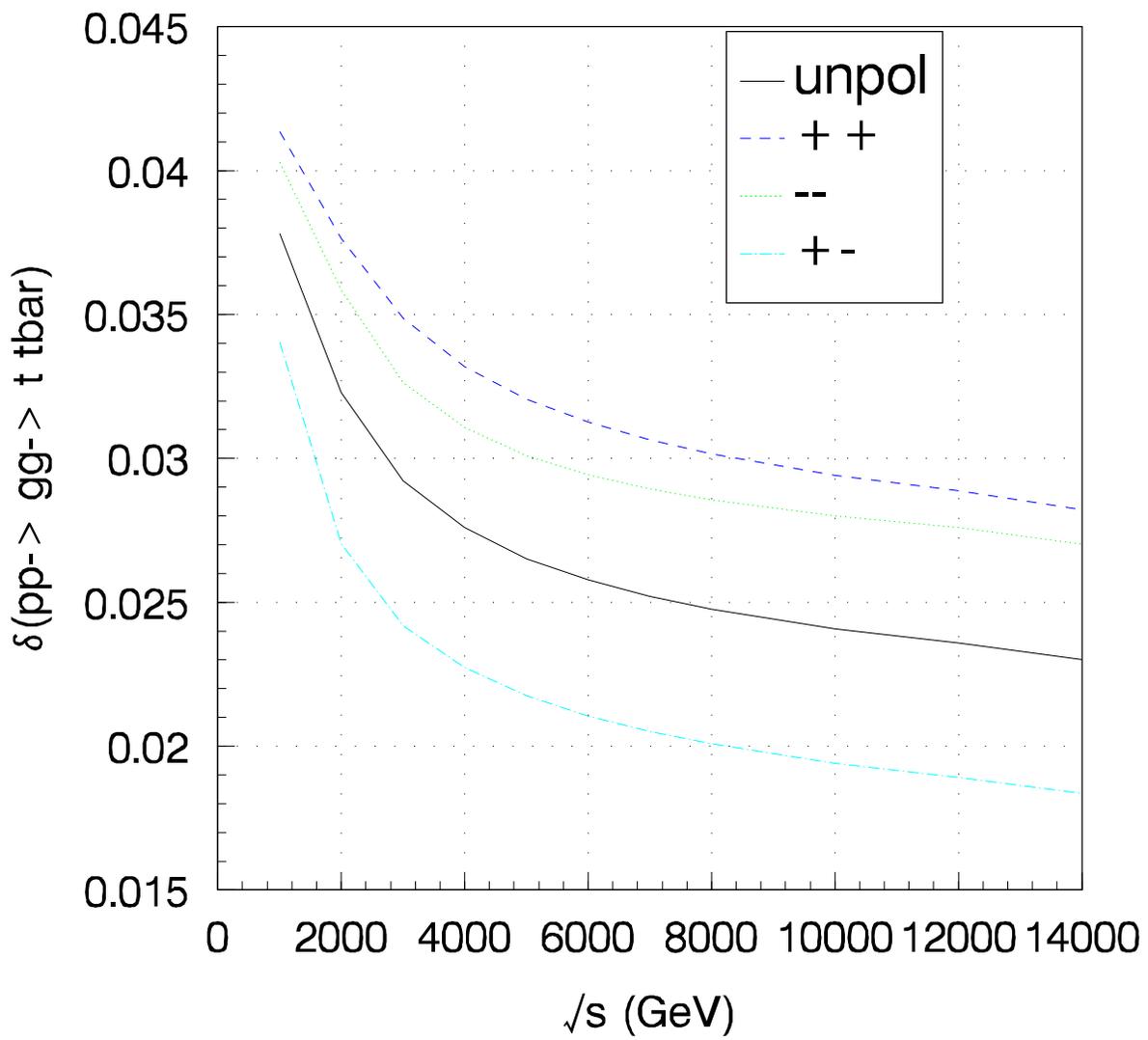
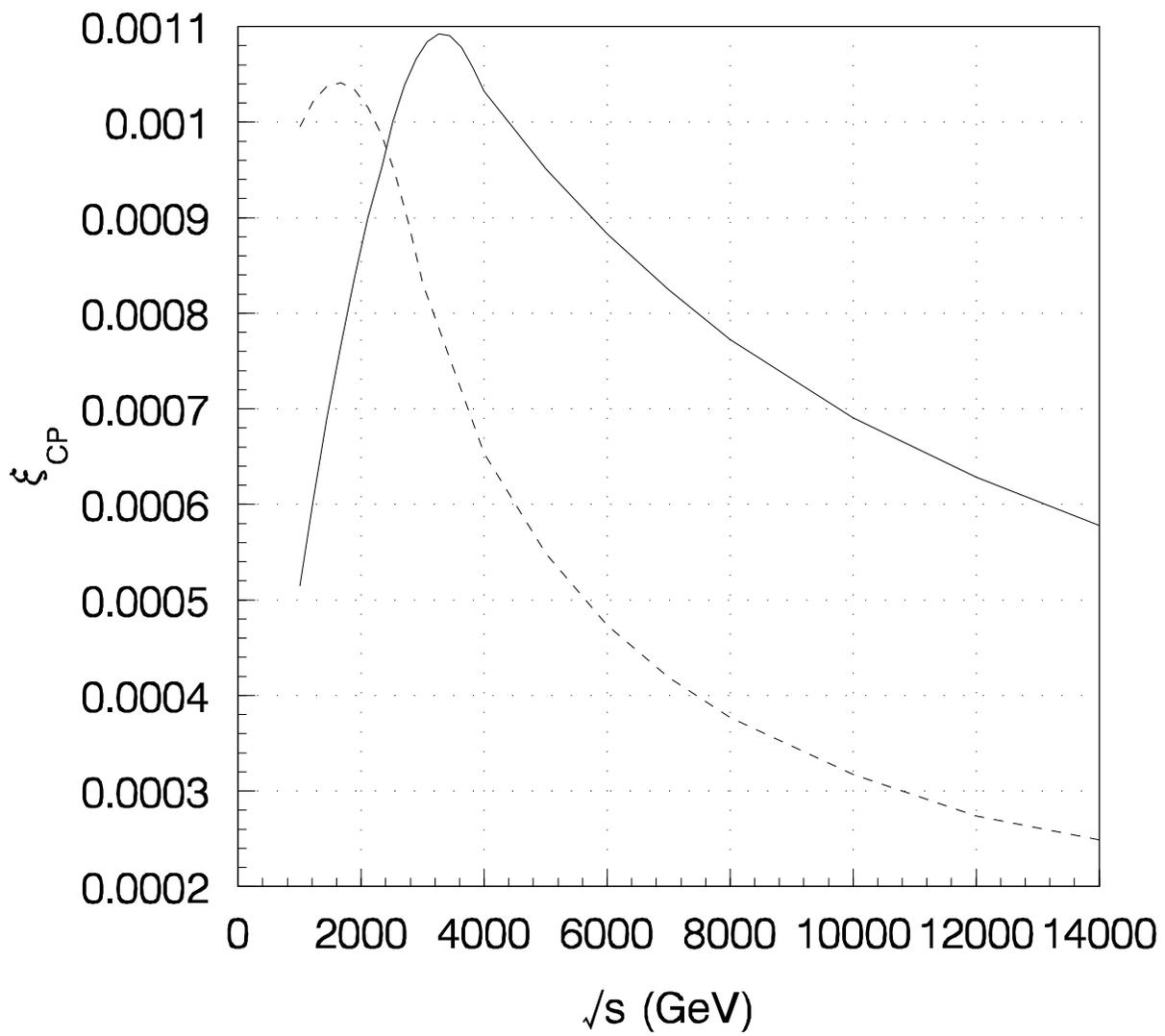
{\bf Fig.2}
    (a) relative corrections to polarized and unpolarized
    cross sections of the $t\bar{t}$ production process in pp colliders
    as a function of $\sqrt{s}$ with input structure functions of Brodsky
    et al. \cite{Geh}(LO); \\
    (b) relative corrections to polarized and unpolarized
    cross sections of the $t\bar{t}$ production process in pp colliders
    as a function of $\sqrt{s}$ with input structure functions of Gl\"uck
    et al. \cite{Glk} \cite{GLK} \cite{STR}(NLO),
    in both above figures,
    solid line for the MSSM QCD correction with unpolarized protons,
    dashed line for the MSSM QCD correction with $proton(+) proton(+)$
    polarization,
    dotted line for the MSSM QCD correction with $proton(-)
    proton(-)$ polarization and
    dot-dashed line for the MSSM QCD correction with $proton(+)
    proton(-)$
    polarization;   \\
    (c) the CP-violating parameter $\xi_{CP}$ as a function of $\sqrt{s}$,
    solid line for input structure functions of Gl\"uck et al(NLO),
    dashed line for input structure functions of Brodsky et al.(LO)

\par
 $m_{\tilde{g}}=200~GeV$, $m_{\tilde{t}_1}=250~GeV$,
$m_{\tilde{t}_2}=450~GeV$ and $\theta=\phi=45^{\circ}$.

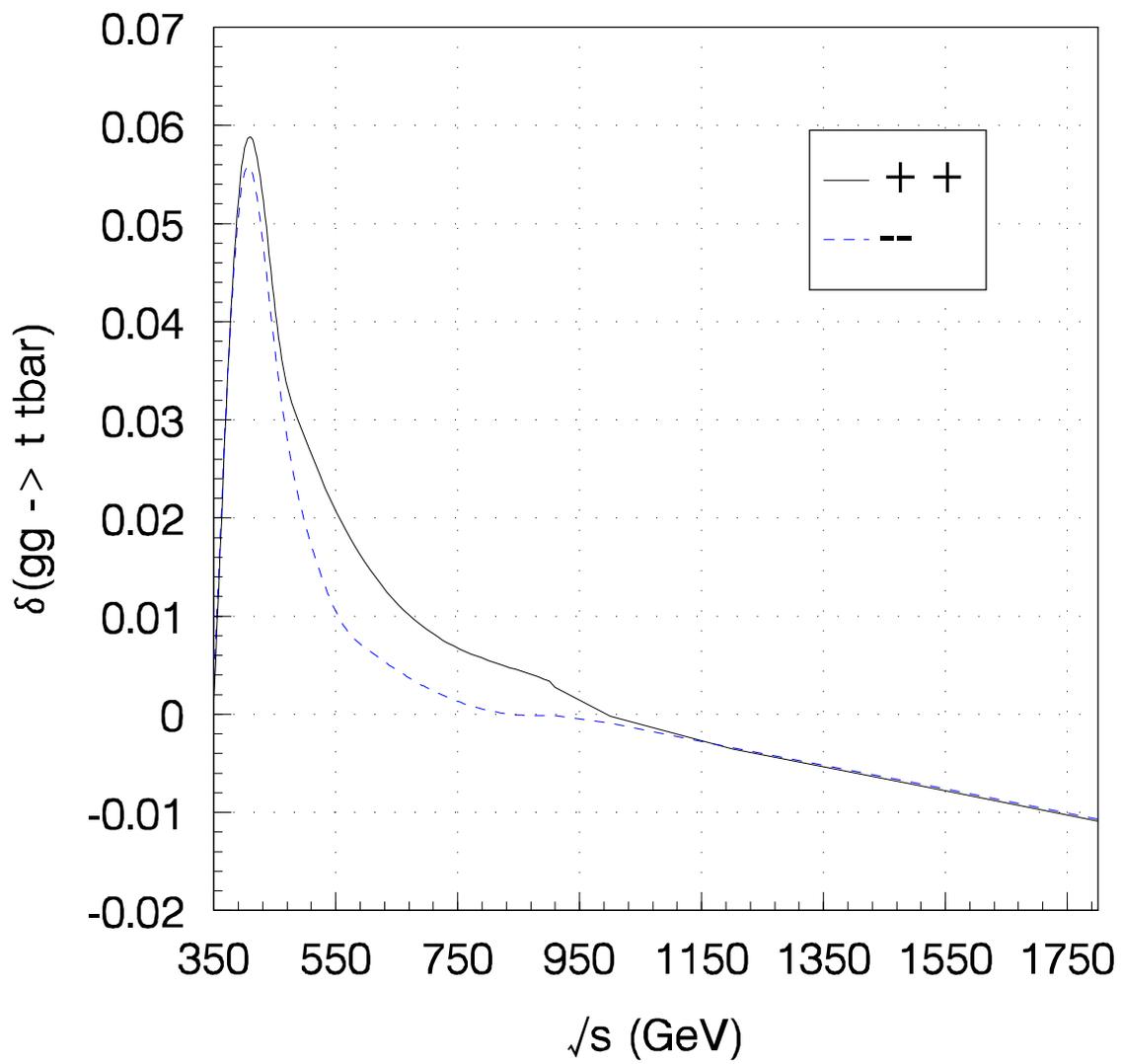
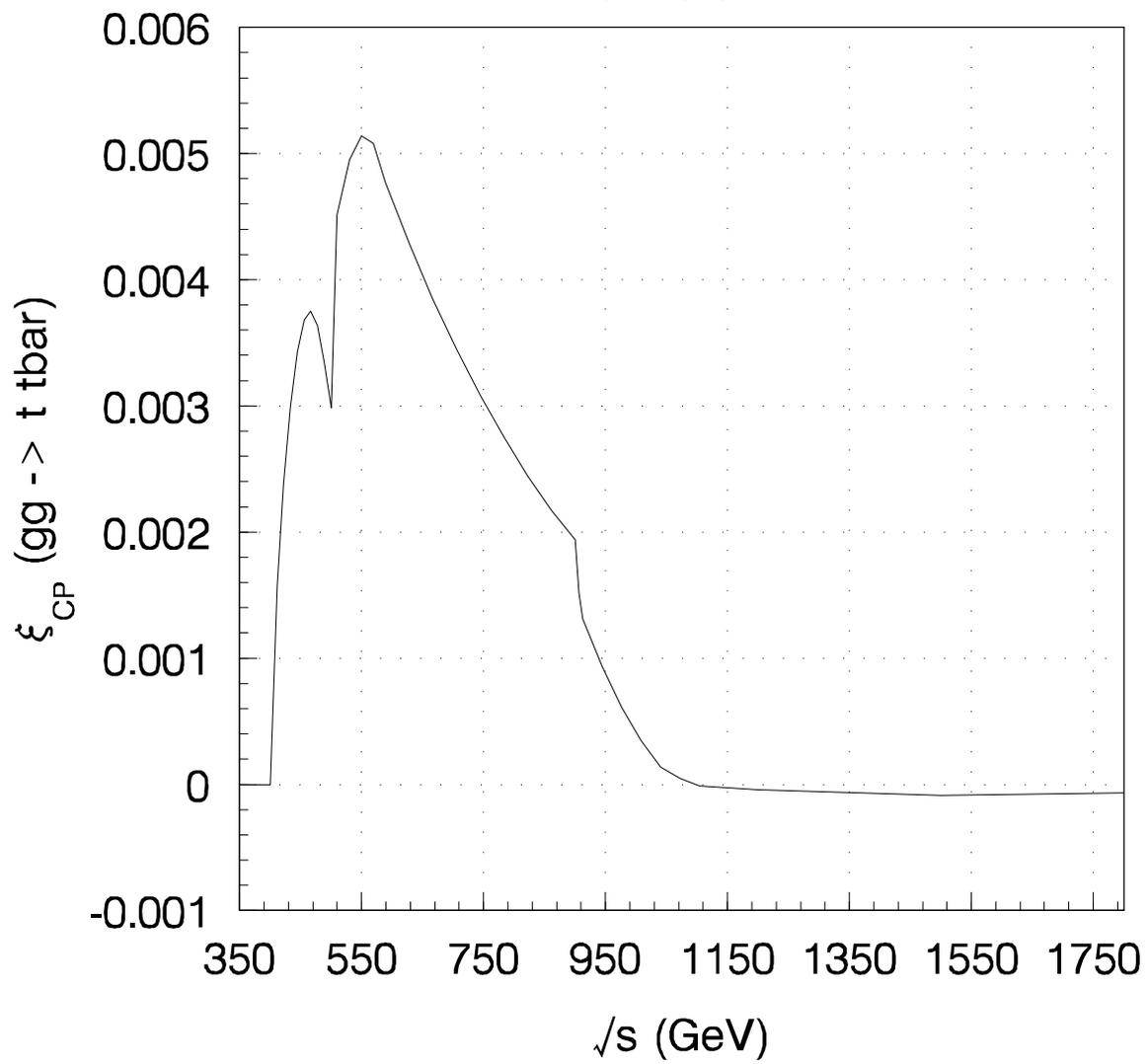
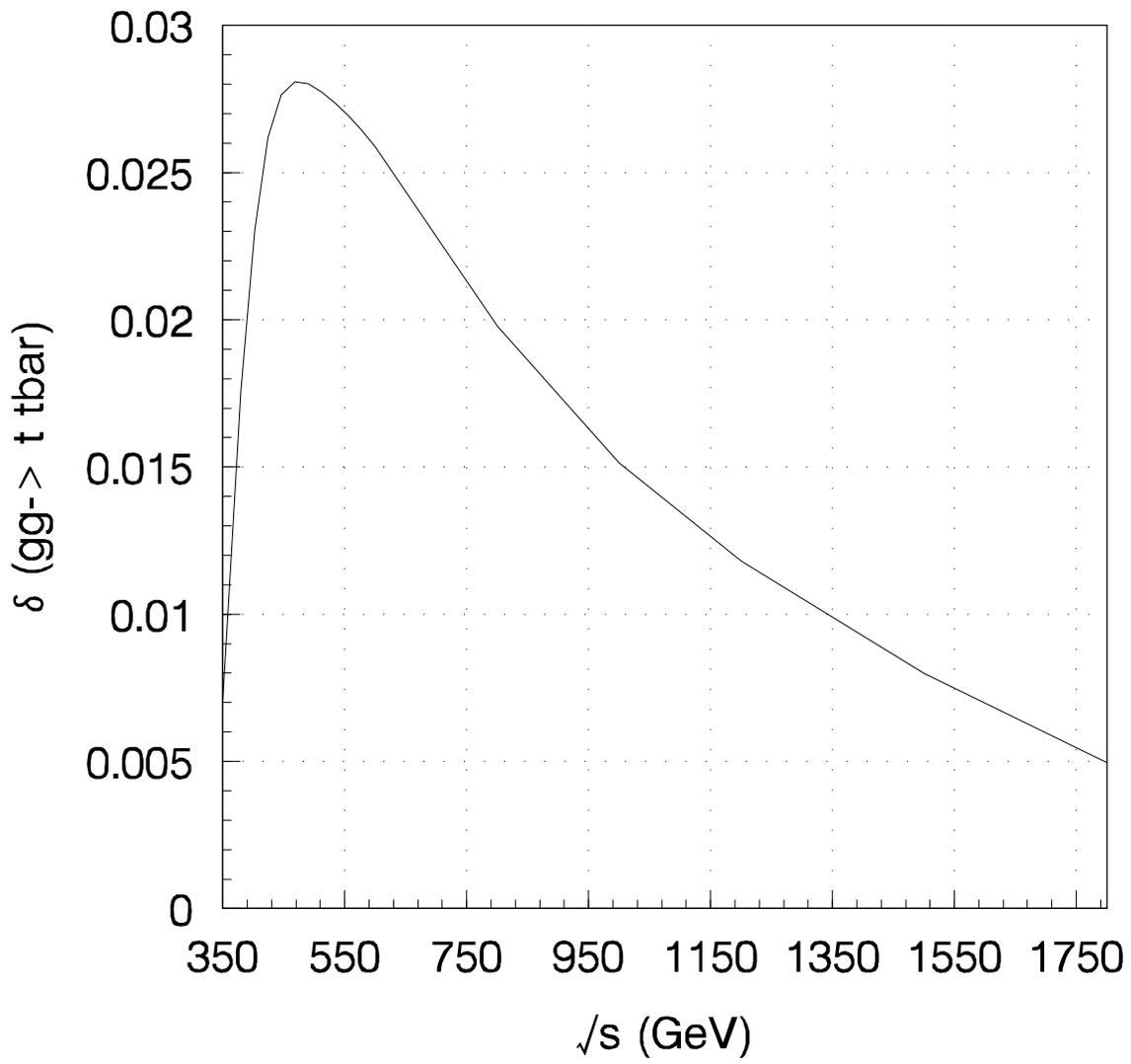
{\bf Fig.3}
 (a) relative corrections to the cross section of the $t\bar{t}$
production
    subprocess, $\hat{\delta}_{\pm \pm}$ as a function of
    $\sqrt{\hat{s}}$,
    solid line for the MSSM QCD correction with $gluon(+) gluon(+)$
    polarization and
    dashed line for the MSSM QCD correction with $gluon(-) gluon(-)$
    polarization.
(b) the CP-violating parameter $\hat{\xi}_{CP}$ of the subprocess as a function
    of $\sqrt{\hat{s}}$.
(c) relative corrections to the cross section of the subprocess
    $\hat{\delta}_{+-}$ as a function of $\sqrt{\hat{s}}$.
\par
$m_{\tilde{g}}=200~GeV$, $m_{\tilde{t}_1}=250~GeV$,
$m_{\tilde{t}_2}=450~GeV$ and $\theta=\phi=45^{\circ}$.

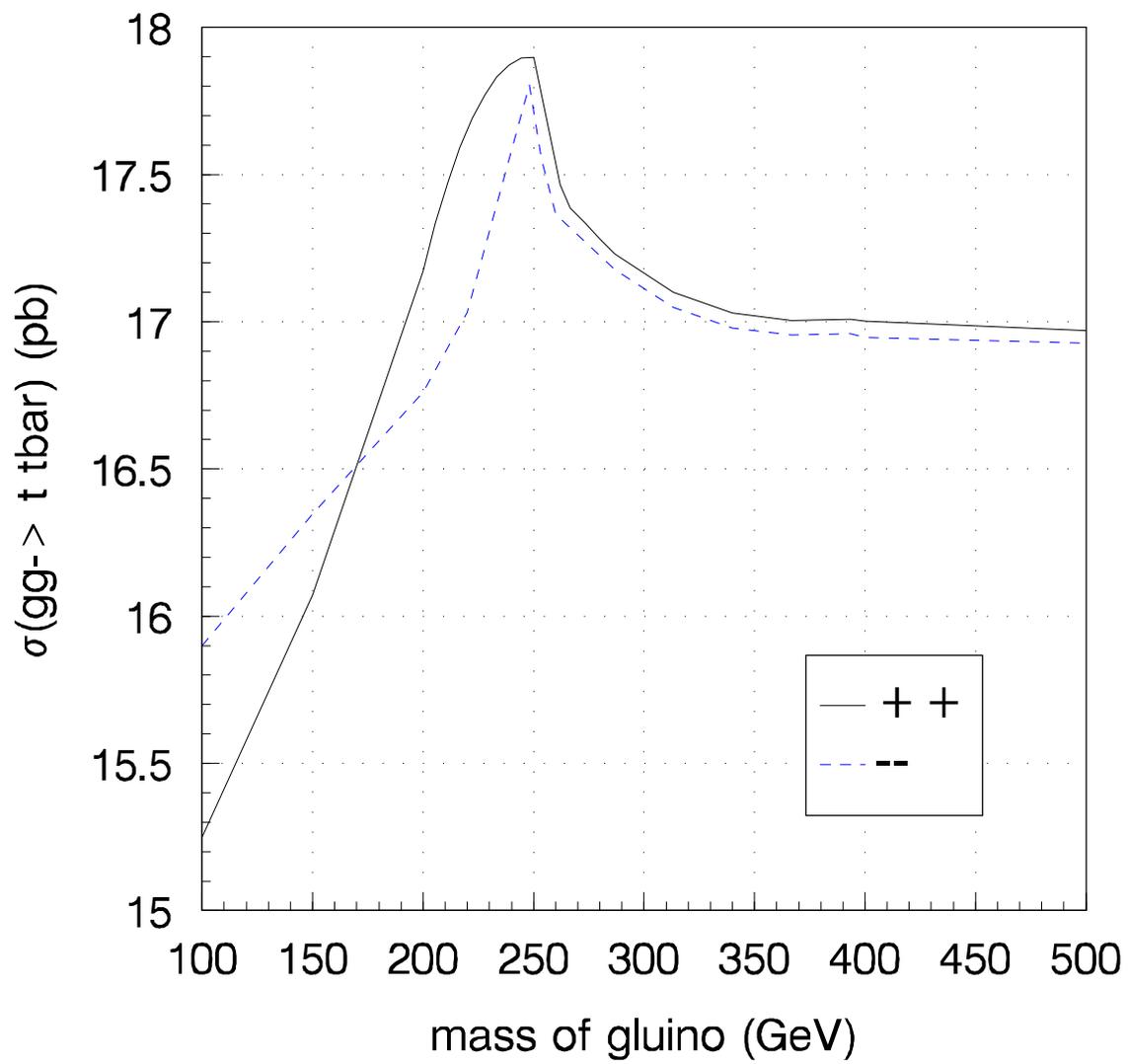
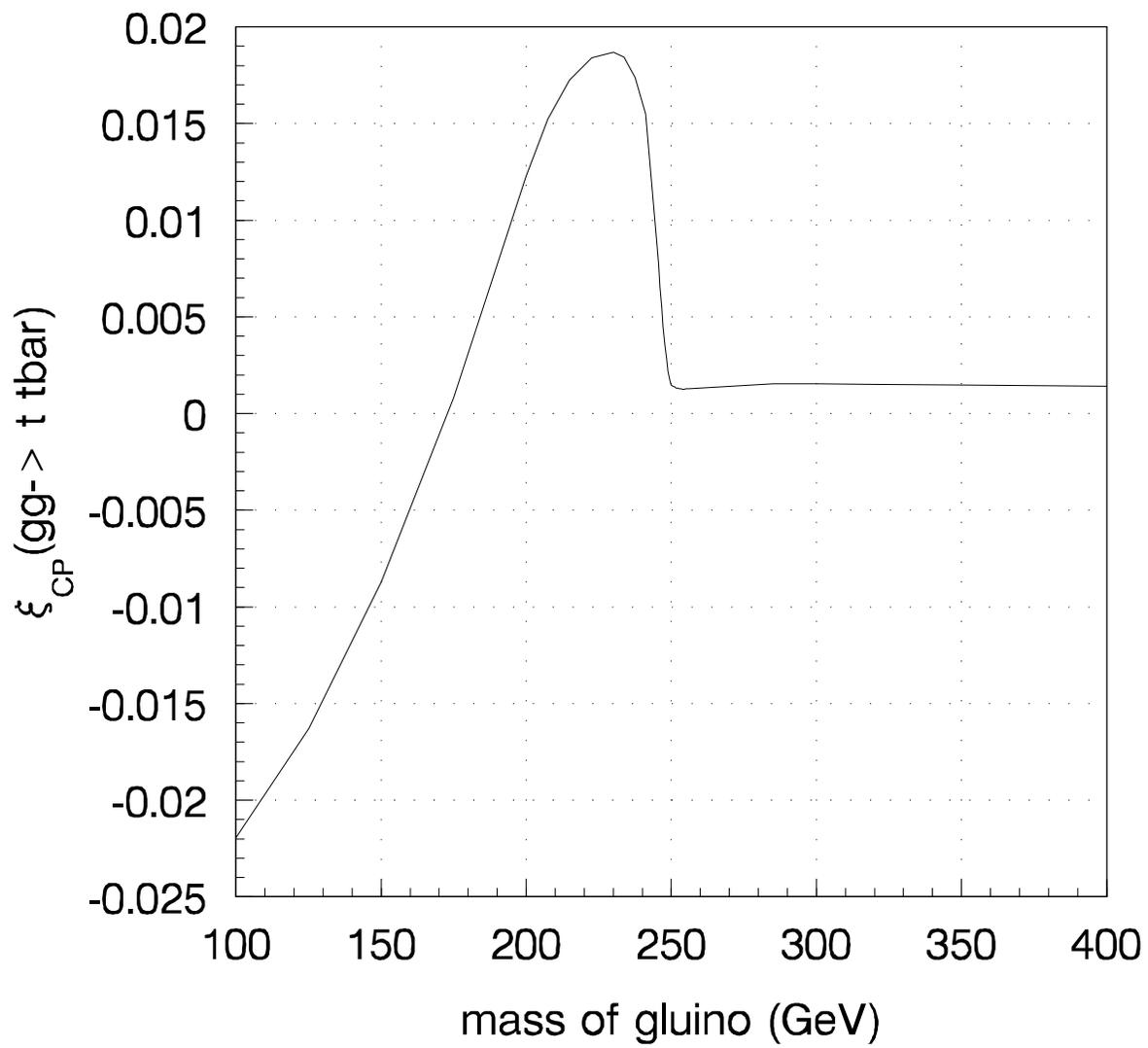
{\bf Fig.4}
  (a) cross section of the $t\bar{t}$ production
    subprocess via gg fusion, $\hat{\sigma}_{\pm \pm}$ as a function
    of $m_{\tilde{g}}$,
    solid line for the MSSM QCD correction with $gluon(+) gluon(+)$
    polarization and
    dashed line for the MSSM QCD correction with $gluon(-) gluon(-)$
    polarization.
(b) the CP-violating parameter $\hat{\xi}_{CP}$ of the subprocess as a function
    of $m_{\tilde{g}}$.
\par
  $m_{\tilde{t}_1}=100~GeV$ , $m_{\tilde{t}_2}=450~GeV$,
$\sqrt{\hat{s}}= 500~GeV$ and $\theta=\phi=45^{\circ}$.

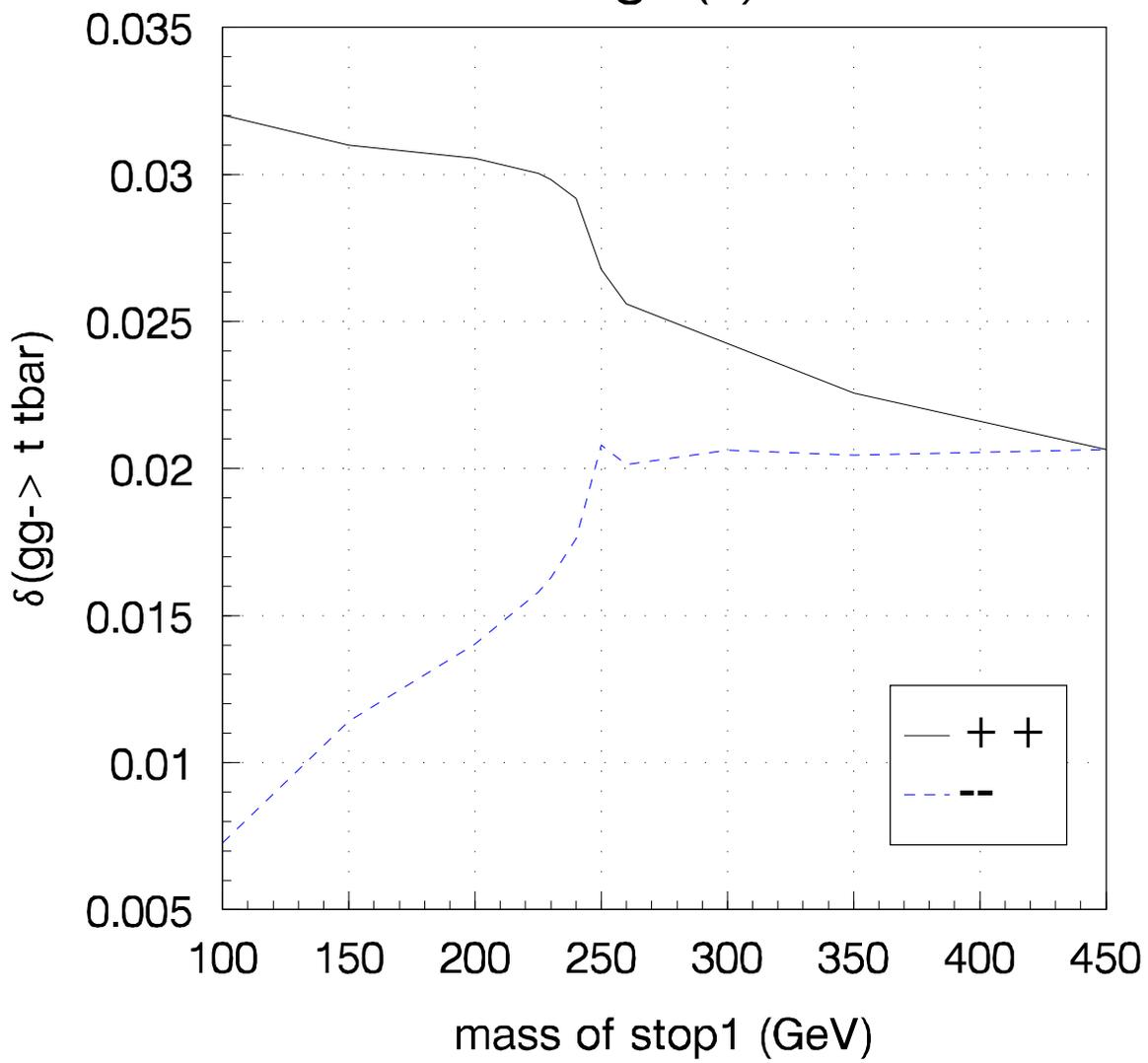
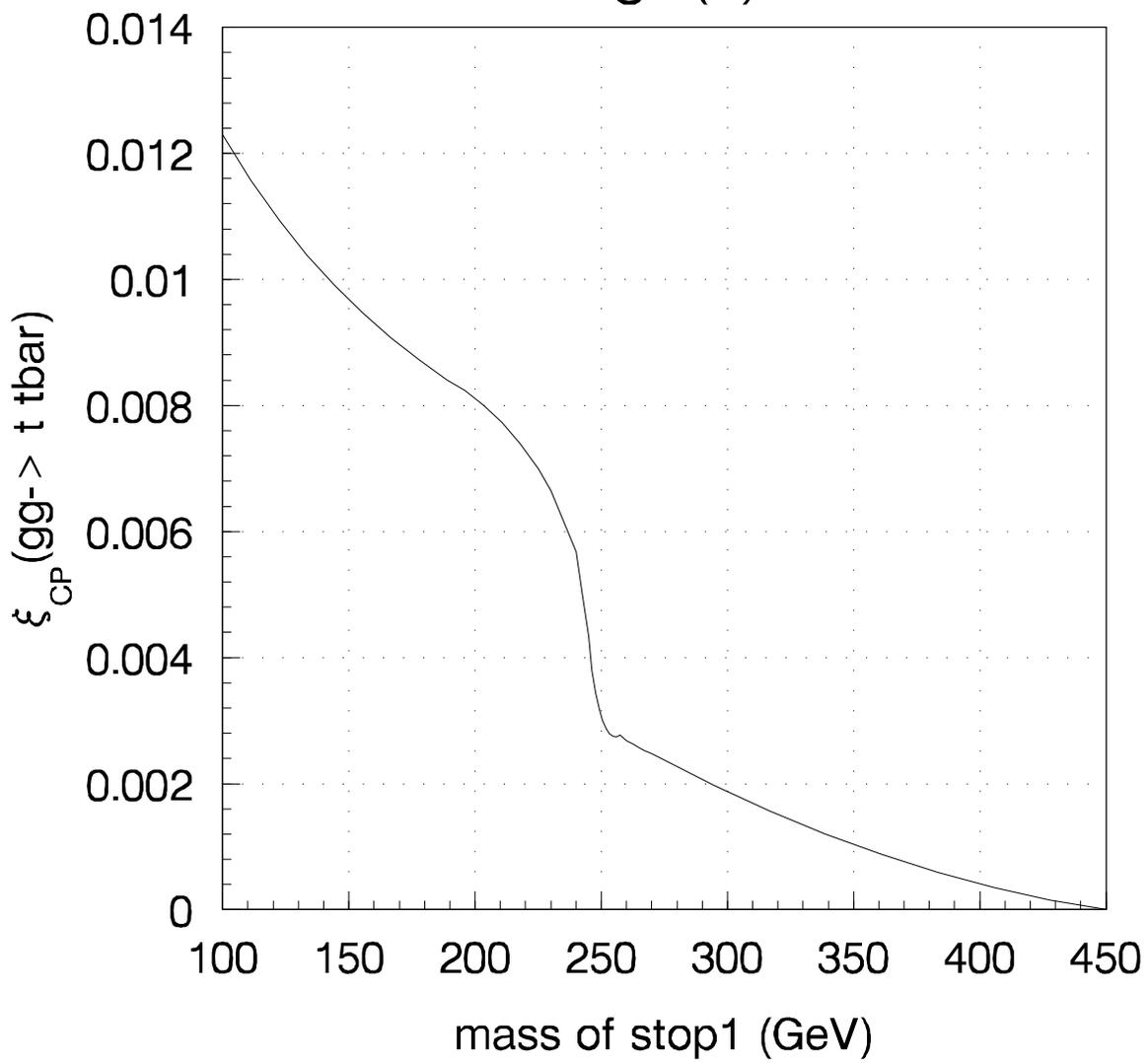
{\bf Fig.5}
(a) relative corrections to the cross section of the $t\bar{t}$ production
    subprocess via gg fusion, $\hat{\delta}_{\pm \pm}$ as a
    function of $m_{\tilde{t}_{1}}$, solid line for the MSSM QCD correction
    with $gluon(+) gluon(+)$ polarization and
    dashed line for the MSSM QCD correction with $gluon(-) gluon(-)$
    polarization.
(b) the CP-violating parameter $\hat{\xi}_{CP}$ of the subprocess as a function
    of $m_{\tilde{t}_{1}}$.
\par
 $m_{\tilde{g}}=200~GeV$, $m_{\tilde{t}_2}=450~GeV$,
$\sqrt{\hat{s}}=500~GeV$ and $\theta=\phi=45^{\circ}$.

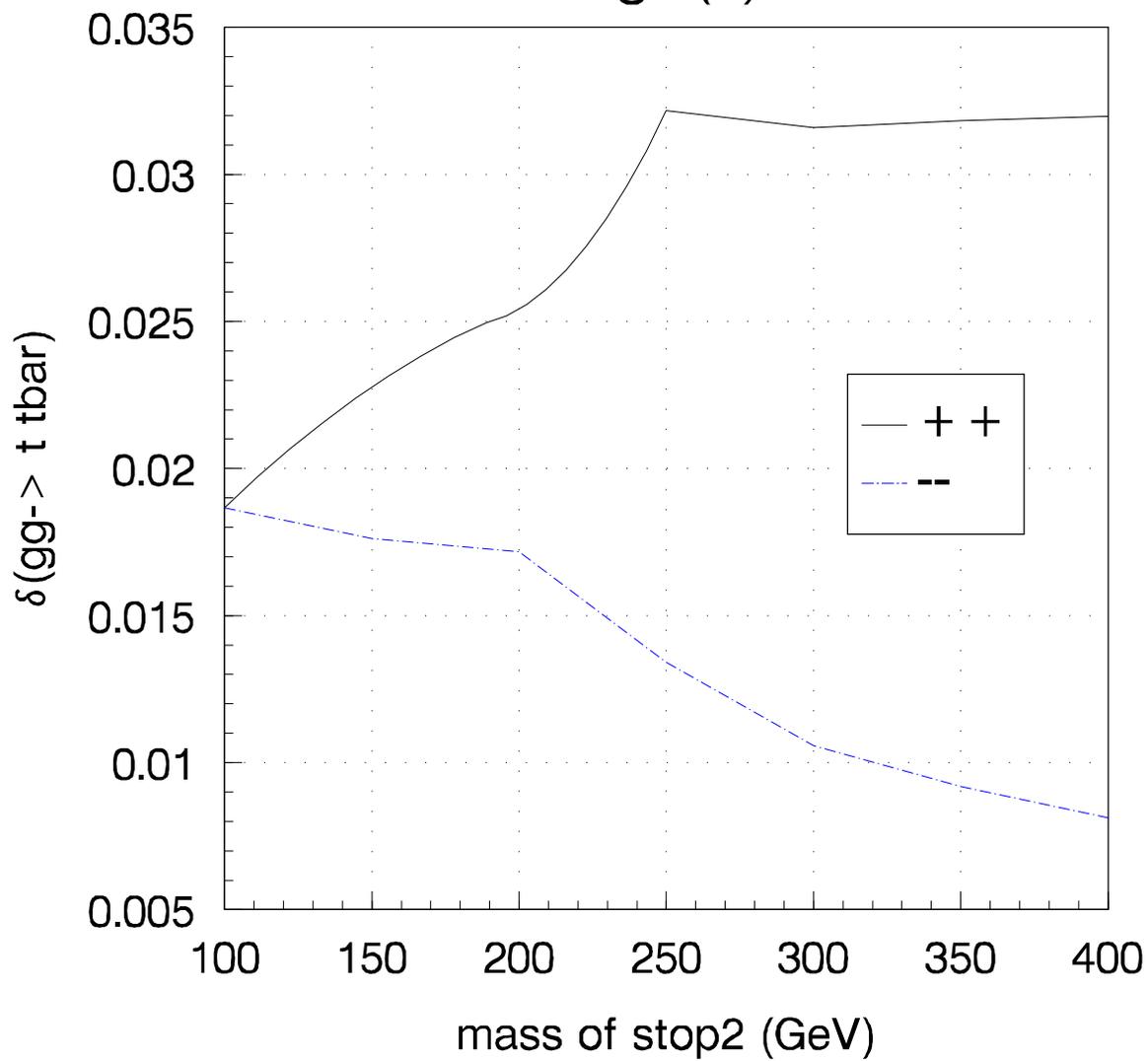
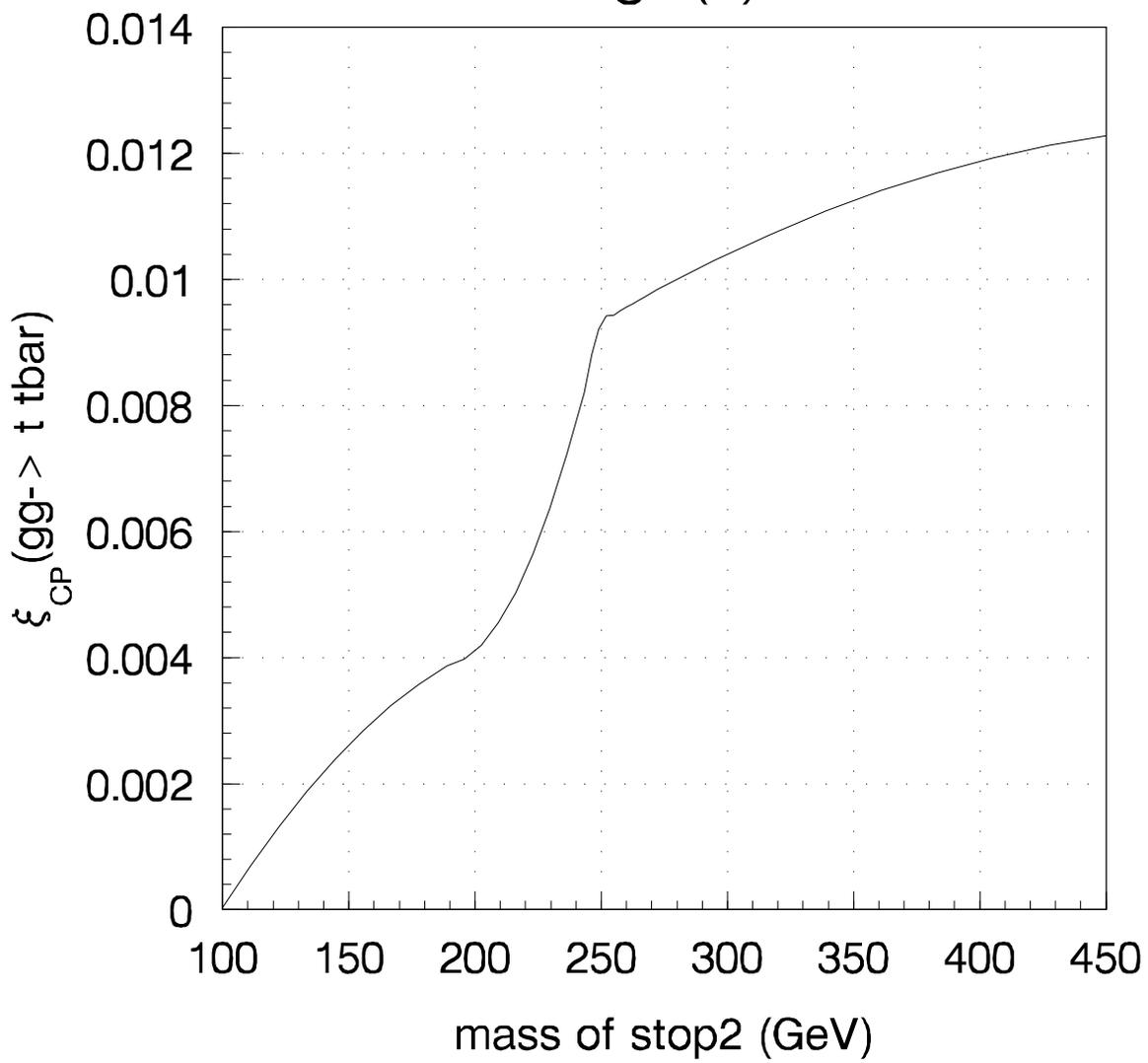
{\bf Fig.6}
(a) relative corrections to the cross section of the $t\bar{t}$ production
    subprocess via gg fusion, $\hat{\delta}_{\pm \pm}$ as a
    function of $m_{\tilde{t}_{2}}$, solid line for the MSSM QCD
    correction with $gluon(+) gluon(+)$ polarization and
    dashed line for the MSSM QCD correction with $gluon(-) gluon(-)$
    polarization.
(b) the CP-violating parameter $\hat{\xi}_{CP}$ of the subprocess as a
    function of $m_{\tilde{t}_{2}}$.
\par
 $m_{\tilde{g}}=200~GeV$, $m_{\tilde{t}_1}=100~GeV$ and
$\sqrt{\hat{s}}=500~GeV$ and $\theta=\phi=45^{\circ}$.

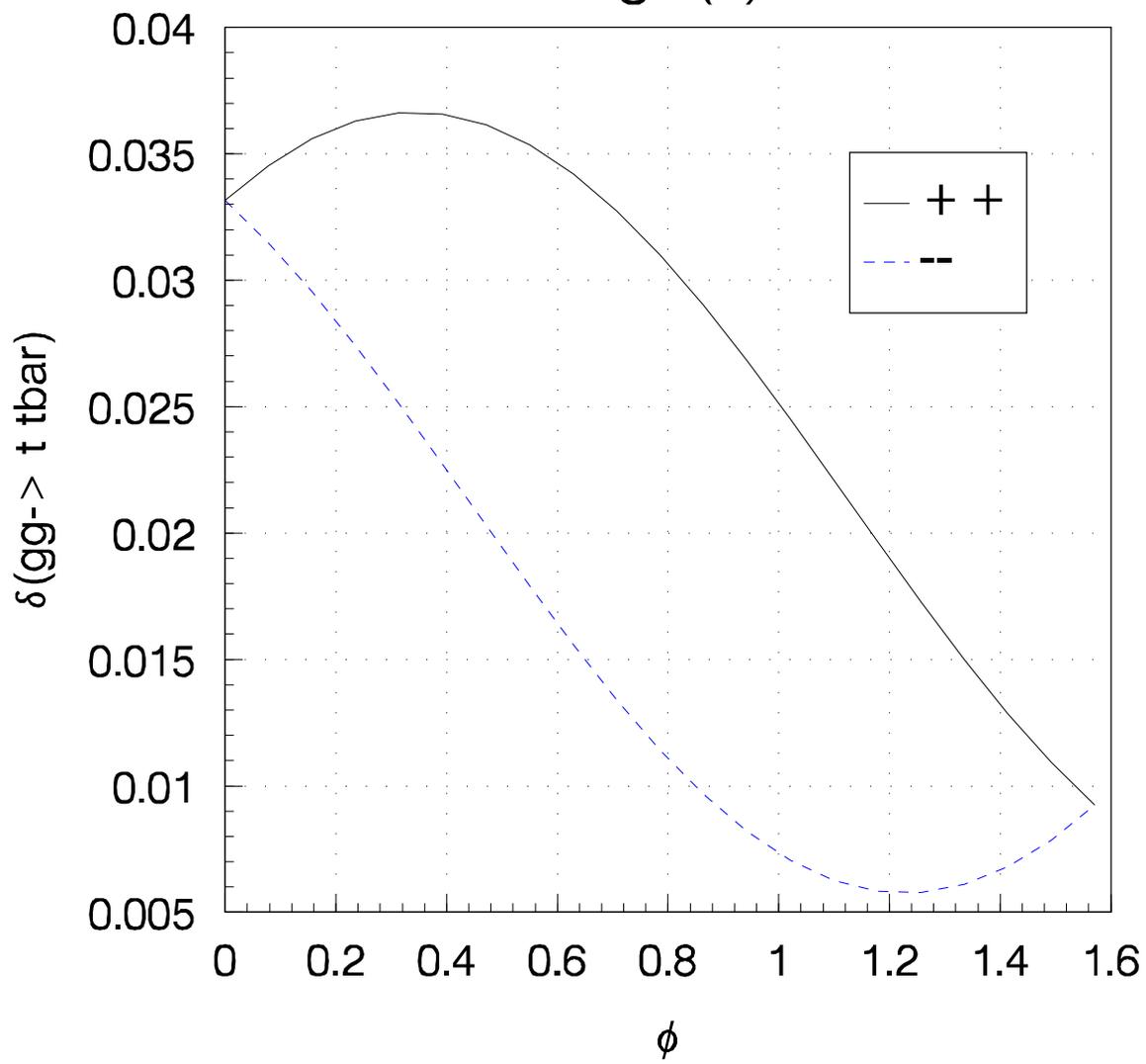
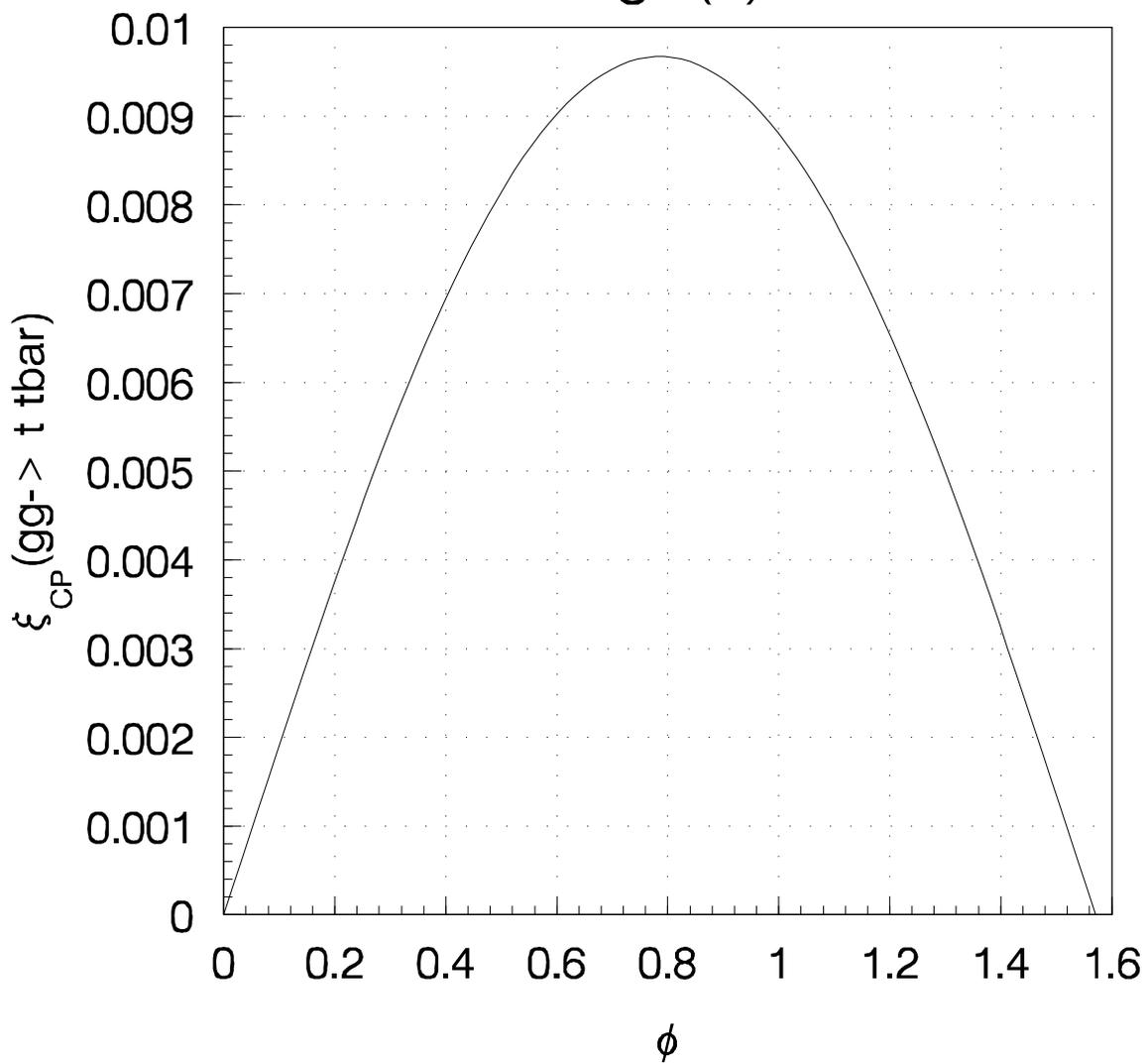
{\bf Fig.7}
(a) relative corrections to the cross section of the $t\bar{t}$ production
    subprocess via gg fusion, $\hat{\delta}_{\pm \pm}$ as a function
    of $\phi$, solid line for the MSSM QCD correction with
    $gluon(+) gluon(+)$ polarization and
    dashed line for the MSSM QCD correction with $gluon(-) gluon(-)$
    polarization.
(b) the CP-violating parameter $\hat{\xi}_{CP}$ of the subprocess as a function
     of $\phi$.
\par
 $m_{\tilde{g}}=200 GeV$,$m_{\tilde{t}_1}=150~GeV$,
$m_{\tilde{t}_2}=450~GeV$ ,$\sqrt{\hat{s}}=500~GeV$ and
$\theta=45^{\circ}$.

\end{document}